\documentclass[preprint,showpacs,preprintnumbers,amsmath,amssymb]{article}
\usepackage{txfonts}

\usepackage{graphicx}
\usepackage{dcolumn}
\usepackage{bm}

\begin{document}

\title{Dynamics of Scalar Field in Polymer-like Representation}

\author{\\ Muxin Han$^{1,2}$\footnote{Email\ address:\ mhan1@lsu.edu}\ \
and
Yongge Ma$^{1}$\footnote{Email\ address:\ mayg@bnu.edu.cn}\\ \\
\small $1.$ Department of Physics, Beijing Normal University, \\
\small Beijing 100875, CHINA\\ \small $2.$ Horace Hearne Jr.
Institute for Theoretical Physics, \\ \small Louisiana State University, \\
\small Baton Rouge, LA 70803, USA}

\date{\today}

\maketitle

\begin{abstract}
In recent twenty years, loop quantum gravity, a background
independent approach to unify general relativity and quantum
mechanics, has been widely investigated. We consider the quantum
dynamics of a real massless scalar field coupled to gravity in this
framework. A Hamiltonian operator for the scalar field can be well
defined in the coupled diffeomorphism invariant Hilbert space, which
is both self-adjoint and positive. On the other hand, the
Hamiltonian constraint operator for the scalar field coupled to
gravity can be well defined in the coupled kinematical Hilbert
space. There are 1-parameter ambiguities due to scalar field in the
construction of both operators. The results heighten our confidence
that there is no divergence within this background independent and
diffeomorphism invariant quantization approach of matter coupled to
gravity. Moreover, to avoid possible quantum anomaly, the master
constraint programme can be carried out in this coupled system by
employing a self-adjoint master constraint operator on the
diffeomorphism invariant Hilbert space.
\end{abstract}


{PACS number(s): 04.60.Pp, 04.60.Ds}

\section{Introduction}
The research on quantum gravity theory is rather active. Many
quantization programmes for gravity are being carried out (for a
summary see e.g. \cite{thiemann2}). In these different kinds of
approaches, the idea of loop quantum gravity is motivated by
researchers in the community of general relativity. It follows
closely the thoughts of general relativity, and hence it is a
quantum theory born with background independency \cite{rovelli}. For
recent review in this field, we refer to
\cite{lecture}\cite{AL}\cite{HHM}.

To apply the background independent quantization technique, one
first casts general relativity into the Hamiltonian formalism of a
diffeomorphism invariant Yang-Mills gauge field theory with a
compact internal gauge group \cite{Ash1}\cite{barbero}. The
kinematical Hilbert space $\mathcal{H}_{kin}^{GR}$ of the quantum
theory is then constructed rigorously. One can even solve the
Gaussian and diffeomorphism constraints to arrive at a
diffeomorphism invariant Hilbert space \cite{ALM}. Certain
geometrical operators are shown to have discrete spectra in the
kinematical Hilbert space
\cite{length}\cite{rovelli8}\cite{area}\cite{volume}\cite{ma2}.
However, some important elements in this approach are not yet
understood. Despite the systematic efforts in constructing the
Hamiltonian constraint operator\cite{thiemann1}\cite{thiemann16}
and the master constraint
operator\cite{thiemann3}\cite{thiemann15}\cite{HM2}, the dynamics
of the quantum theory has not been fully understood, especially if
one wants to include all known matter fields. The primary goal of
this paper is to apply the loop quantization technique to a scalar
field and check whether the quantum dynamics can be well defined.
Since we will use the developed polymer-like kinematical
description of the scalar field\cite{ALS}\cite{KLB}, it can be
considered as a development of the construction for quantum Higgs
fields in Refs. \cite{higgs} and \cite{thiemann7}.

In section 2, the Hamiltonian formalism of a massless real scalar
field coupled to gravity is obtained in generalized Palatini
formulation. For readers' convenience, the loop quantum kinematical
setting of a real scalar field coupled to gravity is also
introduced. We then show in section 3 that an operator corresponding
to the Hamiltonian of the scalar field can be well defined on the
coupled diffeomorphism invariant Hilbert space. It is even positive
and self-adjoint. Thus quantum gravity acts exactly as a natural
regulator for the quantum scalar field in the polymer
representation. In section 4, to study the whole dynamical system of
the scalar field coupled to gravity, a Hamiltonian constraint
operator is defined in the coupled kinematical Hilbert space.
Moreover, the contribution of the scalar field to the Hamiltonian
constraint can be promoted to a positive self-adjoint operator.
Similar to the gravitational Hamiltonian constraint, there is an
one-parameter ambiguity in defining both the Hamiltonian operator
and the constraint operator due to the scalar field. To avoid
possible quantum anomaly and find physical Hilbert space, the
programme of master constraint for the coupled system is discussed
in section 5. A self-adjoint master operator is obtained in the
diffeomorphism invariant Hilbert space, which assures the
feasibility of the programme.

\section{Polymer-like representation for scalar field coupled to gravity}

Consider first the classical dynamics of a real massless scale field
$\phi$ coupled to gravity on a 4-manifold $M$. The coupled
generalized Palatini action reads
\begin{eqnarray}
S[e_{K}^{\beta},\omega_{\alpha}^{\
IJ},\phi]=S_{p}[e_{K}^{\beta},\omega_{\alpha}^{\
IJ}]+S_{KG}[e_{K}^{\beta},\phi],\nonumber
\end{eqnarray}
where
\begin{eqnarray}
S_{p}[e_{K}^{\beta},\omega_{\alpha}^{\ IJ}]
=\frac{1}{2\kappa}\int_{M}d^4x(e)
e_{I}^{\alpha}e_{J}^{\beta}(\Omega_{\alpha\beta}^{\ \
IJ}+\frac{1}{2\gamma}\epsilon^{IJ}_{\ \ KL}\Omega_{\alpha\beta}^{\
\ KL}),\nonumber\\
S_{KG}[e_{K}^{\beta},\phi]=-\frac{\alpha_M}{2}\int_{M}d^4x(e)
\eta^{IJ}e^{\alpha}_{I}e^{\beta}_{J}(\partial_\alpha\phi)\partial_\beta\phi,\nonumber
\end{eqnarray}
here $e_{K}^{\beta}$ and $\omega_{\alpha}^{\ IJ}$ are respectively
the tetrad and Lorentz connection on $M$, the real number $\gamma$,
$k$ and $\alpha_M$ are respectively the Barbero-Immirzi parameter,
the gravitational constant and the coupling constant. From now on we
use $\alpha,\ \beta, \cdots$ for 4-dimensional spacetime indices and
$I,\ J, \cdots$ for internal Lorentz indices, $a,\ b, \cdots$ for
3-dimensional spacial indices and $i,\ j, \cdots$ for internal
$SU(2)$ indices. After 3+1 decomposition and Legendre
transformation, similar to the case in Palatini formalism\cite{han},
we obtain the total Hamiltonian of the coupling system on the
3-manifold $\Sigma$ as:
\begin{eqnarray}
\mathcal{H}_{tot}=\int_\Sigma(\Lambda^iG_i+N^a\mathcal{V}_a+NC),\label{hamilton}
\end{eqnarray}
where $\Lambda^i$, $N^a$ and $N$ are Lagrange multipliers, and the
Gaussian, diffeomorphism and Hamiltonian constraints are expressed
respectively as:
\begin{eqnarray}
G_i&=&D_a\widetilde{P}^a_i\ :=\
\partial_a\widetilde{P}^a_i+\epsilon_{ij}^{\
\ k}A_a^j\widetilde{P}^a_k,\\
\mathcal{V}_a&=&\widetilde{P}^b_iF_{ab}^i-A^i_aG_i+\widetilde{\pi}\partial_a\phi,\\
C&=&\frac{\kappa\gamma^2}{2\sqrt{|\det
q|}}\widetilde{P}^a_i\widetilde{P}^b_j[\epsilon^{ij}_{\ \
k}F^k_{ab}-2(1+\gamma^2)K^i_{[a}K^j_{b]}]\nonumber\\
&+&\frac{1}{\sqrt{|\det
q|}}[\frac{\kappa^2\gamma^2\alpha_M}{2}\delta^{ij}\widetilde{P}^{a}_{i}\widetilde{P}^{b}_{j}(\partial_a
\phi)\partial_b\phi
+\frac{1}{2\alpha_M}\widetilde{\pi}^2],\label{kgconstraint}
\end{eqnarray}
here the conjugate pair for gravity consists of the $SU(2)$
connection $A^i_a$ and the densitized triad $\widetilde{P}^b_j$,
$F_{ab}^i$ is the curvature of $A^i_a$, and $\widetilde{\pi}$
denotes the momentum conjugate to $\phi$. Thus one has the
elementary Poisson brackets
\begin{eqnarray}
\{A^i_a(x),\widetilde{P}^b_j(y)\}&=&\delta^a_b\delta^i_j\delta(x-y),\nonumber\\
\{\phi(x),\widetilde{\pi}(y)\}&=&\delta(x-y).\nonumber
\end{eqnarray}
Note that the second term of the Hamiltonian constraint
(\ref{kgconstraint}) is just the Hamiltonian of the real scalar
field.

Now we introduce the background independent quantization of a real
scalar field coupled to gravity, following the polymer
representation of the scalar field \cite{ALS}\cite{KLB}. The
classical configuration space, $\mathcal{U}$, consists of all
real-valued smooth functions $\phi$ on $\Sigma$. Given a set of
finite number of points $X=\{x_1,...,x_N\}$ in $\Sigma$, denote
$Cyl_X$ the vector space generated by finite linear combinations of
the following functions of $\phi$:
\begin{eqnarray}
\Pi_{X,\bf{\lambda}}(\phi):=\prod_{x_j\in X}\exp[i
\lambda_j\phi(x_j)],\nonumber
\end{eqnarray}
where $\mathbf{\lambda}\equiv (\lambda_1, \lambda_2,
\cdot\cdot\cdot, \lambda_N)$ are arbitrary real numbers. It is
obvious that $Cyl_X$ has the structure of a $*$-algebra. The space
$Cyl$ of all cylindrical functions on $\mathcal{U}$ is defined by
\begin{equation}
Cyl:=\cup_X Cyl_X.
\end{equation}
Completing $Cyl$ with respect to the sup norm, one obtains a unital
Abelian $C$*-algebra $\overline{Cyl}$. Thus one can use the GNS
structure to construct its cyclic representations. A preferred
positive linear functional $\omega_0$ on $\overline{Cyl}$ is defined
by
\begin{eqnarray}
\omega_0(\Pi_{X,\bf{\lambda}})=\left\{%
\begin{array}{ll}
    1 & \hbox{if $\lambda_j=0$ $\forall j$} \\
    0 & \hbox{otherwise}, \\
\end{array}%
\right.\ \nonumber
\end{eqnarray}
which defines a diffeomorphism-invariant faithful Borel measure
$\mu$ on $\mathcal{U}$ as
\begin{eqnarray}\label{smeasure}
\int_\mathcal{U}d\mu(\Pi_{X,\bf{\lambda}})=\left\{%
\begin{array}{ll}
    1 & \hbox{if $\lambda_j=0$ $\forall j$} \\
    0 & \hbox{otherwise}. \\
\end{array}%
\right.\
\end{eqnarray}
Thus one obtains the Hilbert space, $\mathcal{H}^{KG}_{kin}\equiv
L^2(\overline{\mathcal{U}}, d\mu)$, of square integrable functions
on a compact topological space $\overline{\mathcal{U}}$ with respect
to $\mu$, where $\overline{Cyl}$ acts by multiplication. The quantum
configuration space $\overline{\mathcal{U}}$ is called the Gel'fand
spectrum of $\overline{Cyl}$. More concretely, for a single point
set $X_0\equiv \{x_0\}$, $Cyl_{X_0}$ is the space of all almost
periodic functions on a real line $\mathbf{R}$. The Gel'fand
spectrum of the corresponding $C$*-algebra $\overline{Cyl}_{X_0}$ is
the Bohr completion $\overline{\mathbf{R}}_{x_0}$ of $\mathbf{R}$
\cite{ALS}, which is a compact topological space such that
$\overline{Cyl}_{X_0}$ is the $C$*-algebra of all continuous
functions on $\overline{\mathbf{R}}_{x_0}$. Since $\mathbf{R}$ is
densely embedded in $\overline{\mathbf{R}}_{x_0}$,
$\overline{\mathbf{R}}_{x_0}$ can be regarded as a completion of
$\mathbf{R}$.

Given a pair $(x_0, \lambda_0)$, there is an elementary
configuration for the scalar field, the so-called point holonomy,
\begin{eqnarray}
U(x_0,\lambda_0):=\exp[i\lambda_0\phi(x_0)].\nonumber
\end{eqnarray}
It corresponds to a configuration operator $\hat{U}(x_0,\lambda_0)$,
which acts on any cylindrical function $\psi(\phi)\in
\mathcal{H}_{kin}^{KG}$ by
\begin{equation}
\hat{U}(x_0,\lambda_0)\psi(\phi)=U(x_0,\lambda_0)\psi(\phi).\nonumber
\end{equation}
All these operators are unitary. But since the family of operators
$\hat{U}(x_0,\lambda)$ fails to be weakly continuous in $\lambda$,
there is no operator $\hat{\phi}(x)$ on $\mathcal{H}^{KG}_{kin}$.
The momentum functional smeared on a 3-dimensional region
$R\subset\Sigma$ is expressed by
\begin{eqnarray}
\pi(R):=\int_R d^3x \widetilde{\pi}(x).\nonumber
\end{eqnarray}
The Poisson bracket between the momentum functional and a point
holonomy can be easily calculated to be
\begin{eqnarray}
\{\pi(R), U(x,\lambda)\}=-i\lambda\chi_R(x)U(x, \lambda),\nonumber
\end{eqnarray}
where $\chi_R(x)$ is the characteristic function for the region $R$.
So the momentum operator is defined by the action on scalar network
functions $\Pi_{c=(X,\bf{\lambda})}$ as
\begin{eqnarray}
\hat{\pi}(R)\ \Pi_{c}(\phi):=i\hbar\{\pi(R),
\Pi_c(\phi)\}=\hbar[\sum_{x_j\in
X}\lambda_j\chi_{R}(x_j)]\Pi_c(\phi).\nonumber
\end{eqnarray}
It is clear from Eq.(\ref{smeasure}) that an orthonormal basis in
$\mathcal{H}^{KG}_{kin}$ is given by the so-called scalar network
functions $\Pi_c(\phi)$, where $c$ denotes $(X(c),\mathbf{\lambda})$
and $\mathbf{\lambda}\equiv (\lambda_1, \lambda_2, \cdot\cdot\cdot,
\lambda_N)$ are non-zero real numbers now. So the total kinematical
Hilbert space $\mathcal{H}_{kin}$ is the direct product of the
kinematical Hilbert space $\mathcal{H}_{kin}^{GR}$ for gravity and
the kinematical Hilbert space for real scalar field, i.e.,
$\mathcal{H}_{kin}:=\mathcal{H}_{kin}^{GR}\otimes\mathcal{H}_{kin}^{KG}$.
Let $\Pi_s(A)$ be the spin network basis in $\mathcal{H}_{kin}^{GR}$
labeled by $s$ \cite{area}\cite{rovelli9}. Then the state
$\Pi_{s,c}\equiv\Pi_s(A)\otimes\Pi_c(\phi)\in
Cyl_{\gamma(s)}(\overline{\mathcal{A}})\otimes Cyl_{X(c)}\equiv
Cyl_{\gamma(s,c)}(\overline{\mathcal{A}})\otimes
Cyl_{\gamma(s,c)}(\overline{\mathcal{U}})$ is a gravity-scalar
cylindrical function on graph $\gamma(s,c)\equiv\gamma(s)\cup X(c)$.
Note that generally $X(c)$ may not coincide with the vertices of the
graph $\gamma(s)$. It is straightforward to see that all of these
functions constitutes a complete set of orthonormal basis in
$\mathcal{H}_{kin}$ as
\begin{eqnarray}
<\Pi_{s',c'}(A,\phi)|\Pi_{s,c}(A,\phi)>_{kin}=\delta_{s's}\delta_{c'c}\
.\nonumber
\end{eqnarray}
Note that none of $\mathcal{H}_{kin}$, $\mathcal{H}^{GR}_{kin}$ and
$\mathcal{H}^{KG}_{kin}$ is a separable Hilbert space.

Now we can consider the quantum dynamics and impose the quantum
constraints on $\mathcal{H}_{kin}$. Firstly, the Gaussian
constraint can be solved independently of
$\mathcal{H}_{kin}^{KG}$, since it only involves gravitational
field. It is also expected that the diffeomorphism constraint can
be implemented by the group averaging strategy in the similar way
as that in the case of pure gravity. Given a spatial
diffeomorphism transformation $\varphi$, a unitary transformation
$\hat{U}_\varphi$ is induced by $\varphi$ in the Hilbert space
$\mathcal{H}_{kin}$, which is expressed as
\begin{eqnarray}
\hat{U}_\varphi\Pi_{s=(\gamma(s),\mathbf{j},\mathbf{N}),c=(X(c),\mathbf{\lambda})}
=\Pi_{\varphi\circ
s=(\varphi(\gamma(s)),\mathbf{j},\mathbf{N}),\varphi\circ
c=(\varphi(X(c)),\mathbf{\lambda})}.\nonumber
\end{eqnarray}
Then the differomorphism invariant spin-scalar-network functions are
defined by group averaging as
\begin{eqnarray}
\Pi_{[s,c]}:=\frac{1}{n_{\gamma(s,c)}}\sum_{\varphi\in
Diff/Diff_{\gamma(s,c)}}\sum_{\varphi'\in
GS_{\gamma(s,c)}}\hat{U}_{\varphi}\hat{U}_{\varphi'}\Pi_{s,c},
\end{eqnarray}
where $Diff_{\gamma}$ is the set of diffeomorphisms leaving the
colored graph $\gamma$ invariant, $GS_{\gamma}$ denotes the graph
symmetry quotient group $Diff_{\gamma}/TDiff_{\gamma}$ where
$TDiff_{\gamma}$ is the set of the diffeomorphisms which is trivial
on the graph $\gamma$, and $n_\gamma$ is the number of elements in
$GS_\gamma$. Following the standard strategy in quantization of pure
gravity, an inner product can be defined on the vector space spanned
by the diffeomorphism invariant spin-scalar-network functions such
that they form an orthonormal basis as:
\begin{eqnarray}
<\Pi_{[s,c]}|\Pi_{[s',c']}>_{Diff}:=\Pi_{[s,c]}[\Pi_{s',c'\in[s',c']}]=\delta_{[s,c],[s',c']}.
\end{eqnarray}
After the completion procedure, we obtain the expected Hilbert space
of diffeomorphism invariant states for the scalar field coupled to
gravity, which is denoted by $\mathcal{H}_{Diff}$. In the following
sections, we would like to discuss the quantum dynamical properties
of the polymer-like scalar field coupled to gravity.

\section{Diffeormorphism invariant quantum Hamiltonian of scalar field}

In the present section, we first consider the quantum scalar field
on a fluctuating background. A similar idea was considered in
Ref.\cite{thiemann8}, where a Hamiltonian operator with respect to
a U(1) group representation of scalar field is defined on a
kinematical Hilbert space $\mathcal{H}_{kin'}$ of matter coupled
to gravity. Then an effective Hamiltonian operator of the scalar
field can be constructed as a quadratic form via
\begin{eqnarray}
&&<\psi_{matter},\
\hat{H}^{eff}_{matter}(m)\ \psi'_{matter}>^{KG}_{kin'}\nonumber\\
&:=&<\psi_{grav}(m)\otimes\psi_{matter},\ \hat{H}_{matter}\
\psi_{grav}(m)\otimes\psi'_{matter}>_{kin'},
\end{eqnarray}
where $\psi_{grav}(m)\in \mathcal{H}^{GR}_{kin}$ presents a
semiclassical state of gravity approximating some classical
spacetime background $m$ where the quantum scalar field lives.
Thus the effective Hamiltonian operator
$\hat{H}^{eff}_{matter}(m)$ of scalar field contains also the
information of the fluctuating background \emph{metric}. In the
light of this idea, we will construct a Hamiltonian operator
$\hat{\mathcal{H}}_{KG}$ for scalar field in the polymer-like
representation. It turns out that the Hamiltonian operator can be
defined in the Hilbert space $\mathcal{H}_{Diff}$ of
diffeomorphism invariant states for scalar field coupled to
gravity without UV-divergence. So the quantum dynamics of the
scalar field is obtained in a diffeomorphism invariant way, which
is expected in the programme of loop quantum gravity. Thus, here
an effective Hamiltonian operator of the scalar field could be
extracted in $\mathcal{H}_{Diff}$ by defining
$<\Psi_{[m]}(A,\phi),\ \hat{\mathcal{H}}_{KG}\
\Psi_{[m]}(A,\phi)>_{Diff}$ to be its expectation value on
diffeomorphism invariant states $\Psi(\phi)$ of scalar field,
where the diffeomorphism invariant semiclassical state
$\Psi_{[m]}(A)$ represents certain fluctuating \emph{geometry}
with spatial diffeomorphism invariance, and the label $[m]$
denotes the classical geometry approximated by $\Psi_{[m]}(A)$.
Moreover, the quadratic properties of the scalar field Hamiltonian
will provide powerful functional analytic tools in the
quantization procedure, such that the self-adjointness of the
Hamiltonian operator can be proved by the theorem in functional
analysis.

\subsection{Regularization of the Hamiltonian}
In a suitable gauge, the classical Hamiltonian for the massless
scalar field on the 3-manifold $\Sigma$ can be given by
\begin{eqnarray}
\mathcal{H}_{KG}&=&\mathcal{H}_{KG,\phi}+\mathcal{H}_{KG,Kin}\nonumber\\
&=&\int_\Sigma
d^3x[\frac{\kappa^2\gamma^2\alpha_M}{2}\frac{1}{\sqrt{|\det
q|}}\delta^{ij}\widetilde{P}^{a}_{i}\widetilde{P}^{b}_{j}(\partial_a\phi)\partial_b\phi+\frac{1}{2\alpha_M}\frac{1}{\sqrt{|\det
q|}}\widetilde{\pi}^2].\label{scalar hamilton}
\end{eqnarray}
We will employ the following identities:
\begin{eqnarray}
\widetilde{P}^a_i=\frac{1}{2\kappa\gamma}\widetilde{\eta}^{abc}\epsilon_{ijk}e^j_be^k_c\
\ \ \ \mathrm{and} \ \ \ \
e^i_a(x)=\frac{2}{\kappa\gamma}\{A^i_a(x),V_{U_x}\},\nonumber
\end{eqnarray}
where $\widetilde{\eta}^{abc}$ denotes the Levi-Civita tensor
tensity and $V_{U_x}$ is the volume of an arbitrary neighborhood
$U_x$ containing the point $x$. By using the point-splitting
strategy, the regulated version of the Hamiltonian (\ref{scalar
hamilton}) is obtained as:
\begin{eqnarray}
\mathcal{H}_{KG,\phi}
&=&\frac{\kappa^2\gamma^2\alpha_M}{2}\int_\Sigma d^3y\int_\Sigma
d^3x\chi_\epsilon(x-y)\delta^{ij}\times\nonumber\\
&&\frac{1}{\sqrt{V_{U^\epsilon_x}}}\widetilde{P}^{a}_{i}(x)(\partial_a\phi(x))\frac{1}{\sqrt{V_{U^\epsilon_y}}}
\widetilde{P}^{b}_{j}(y)\partial_b\phi(y)\nonumber\\
&=&\frac{32\alpha_M}{\kappa^4\gamma^4}\int_\Sigma d^3y\int_\Sigma
d^3x\chi_\epsilon(x-y)\delta^{ij}\times\nonumber\\
&&\widetilde{\eta}^{aec}(\partial_a\phi(x))\mathrm{Tr}\big(\tau_i\{\mathbf{A}_e(x),V_{U^\epsilon_x}^{3/4}\}
\{\mathbf{A}_c(x),V_{U^\epsilon_x}^{3/4}\}\big)\times\nonumber\\
&&\widetilde{\eta}^{bfd}(\partial_b\phi(y))\mathrm{Tr}\big(\tau_j\{\mathbf{A}_f(y),V_{U^\epsilon_y}^{3/4}\}
\{\mathbf{A}_d(y),V_{U^\epsilon_y}^{3/4}\}\big),\nonumber\\
\mathcal{H}_{KG,Kin} &=&\frac{1}{2\alpha_M}\int_\Sigma
d^3x\widetilde{\pi}(x)\int_\Sigma
d^3x\widetilde{\pi}(y)\times\nonumber\\
&&\int_\Sigma
d^3u\frac{\det(e_a^i(u))}{(V_{U^\epsilon_u})^{3/2}}\int_\Sigma
d^3w\frac{\det(e_a^i(v))}{(V_{U^\epsilon_w})^{3/2}}\chi_\epsilon(x-y)\chi_\epsilon(u-x)\chi_\epsilon(w-y)\nonumber\\
&=&\frac{1}{2\alpha_M}\frac{2^8}{9(\kappa\gamma)^6}\int_\Sigma
d^3x\widetilde{\pi}(x)\int_\Sigma
d^3x\widetilde{\pi}(y)\times\nonumber\\
&&\int_\Sigma d^3u\
\widetilde{\eta}^{abc}\mathrm{Tr}\big(\{\mathbf{A}_a(u),\sqrt{V_{U^\epsilon_u}}\}
\{\mathbf{A}_b(u),\sqrt{V_{U^\epsilon_u}}\}\{\mathbf{A}_c(u),\sqrt{V_{U^\epsilon_u}}\}\big)\times\nonumber\\
&&\int_\Sigma d^3w\
\widetilde{\eta}^{def}\mathrm{Tr}\big(\{\mathbf{A}_d(w),\sqrt{V_{U^\epsilon_w}}\}
\{\mathbf{A}_e(w),\sqrt{V_{U^\epsilon_w}}\}\{\mathbf{A}_f(w),\sqrt{V_{U^\epsilon_w}}\}\big)\times\nonumber\\
&&\chi_\epsilon(x-y)\chi_\epsilon(u-x)\chi_\epsilon(w-y),\nonumber
\end{eqnarray}
where we denote $\mathbf{A}_a\equiv A_a^i\tau_i$,
$\chi_\epsilon(x-y)$ is the characteristic function of a box
containing $x$ with scale $\epsilon$ such that
$\lim_{\epsilon\rightarrow0}\chi_\epsilon(x-y)/{\epsilon^3}=\delta(x-y)$,
and $V_{U^\epsilon_x}$ is the volume of the box. In order to
quantize the Hamiltonian as a well-defined operator in polymer-like
representation, we have to express the classical formula of
$\mathcal{H}_{KG}$ in terms of elementary variables with clear
quantum analogs. This can be realized by introducing a triangulation
$T(\epsilon)$ of $\Sigma$, where the parameter $\epsilon$ describes
how fine the triangulation is. The quantity regulated on the
triangulation is required to have correct limit when
$\epsilon\rightarrow0$. Given a tetrahedron $\Delta\in T(\epsilon)$,
we use $\{s_i(\Delta)\}_{i=1,2,3}$ to denote the three outgoing
oriented segments in $\Delta$ with a common beginning point
$v(\Delta)=s(s_i(\Delta))$ and use $a_{ij}(\Delta)$ to denote the
arcs connecting the end points of $s_i(\Delta)$ and $s_j(\Delta)$.
Then several loops $\alpha_{ij}(\Delta)$ are formed by
$\alpha_{ij}(\Delta):=s_i(\Delta)\circ a_{ij}(\Delta)\circ
s_j(\Delta)^{-1}$. Thus we have the identities:
\begin{eqnarray}
\{\int_{s(\Delta)}dt\
\mathbf{A}_a\dot{s}^a(t),V_{U^\epsilon_{s(s(\Delta))}}^{3/4}\}&=&-A(s(\Delta))^{-1}\{A(s(\Delta)),V_{U^\epsilon_{s(s(\Delta))}}^{3/4}\}
+o(\epsilon),\nonumber
\end{eqnarray}
and
\begin{eqnarray}
\int_{s(\Delta)} dt\ \partial_a\phi\dot{s}^a(t)&=&\frac{1}{i\lambda}
U(s(s(\Delta)),\lambda)^{-1}[U(t(s(\Delta)),\lambda)-U(s(s(\Delta)),\lambda)]+o(\epsilon)
\nonumber
\end{eqnarray}
for nonzero $\lambda$, where $s(s(\Delta))$ and $t(s(\Delta))$
denote respectively the beginning and end points of segment
$s(\Delta)$ with scale $\epsilon$ associated with a tetrahedron
$\Delta$. Regulated on the triangulation, the classical Hamiltonian
of scalar field reads
\begin{eqnarray}
\mathcal{H}^\epsilon_{KG,\phi}
&=&-\frac{4\alpha_M}{9\kappa^4\gamma^4}\sum_{\Delta'\in
T(\epsilon)}\sum_{\Delta\in
T(\epsilon)}\chi_\epsilon(v(\Delta)-v(\Delta'))
\delta^{ij}\times\nonumber\\
&&\epsilon^{lmn}\frac{1}{\lambda}U(v(\Delta),\lambda)^{-1}[U(t(s_l(\Delta)),\lambda)-U(v(\Delta),\lambda)]\times\nonumber\\
&&\mathrm{Tr}\big(\tau_iA(s_m(\Delta))^{-1}\{A(s_m(\Delta)),V_{U^\epsilon_{v(\Delta)}}^{3/4}\}A(s_n(\Delta))^{-1}
\{A(s_n(\Delta)),V_{U^\epsilon_{v(\Delta)}}^{3/4}\}\big)\times\nonumber\\
&&\epsilon^{kpq}\frac{1}{\lambda}U(v(\Delta'),\lambda)^{-1}[U(t(s_k(\Delta')),\lambda)-U(v(\Delta'),\lambda)]\times\nonumber\\
&&\mathrm{Tr}\big(\tau_jA(s_p(\Delta'))^{-1}\{A(s_p(\Delta')),V_{U^\epsilon_{v(\Delta')}}^{3/4}\}A(s_q(\Delta'))^{-1}
\{A(s_q(\Delta')),V_{U^\epsilon_{v(\Delta')}}^{3/4}\}\big),\nonumber\\
\mathcal{H}^\epsilon_{KG,Kin}
&=&\frac{16}{81\alpha_M(\kappa\gamma)^6}\sum_{\Delta\in
T(\epsilon)}\sum_{\Delta'\in
T(\epsilon)}{\pi}(\Delta){\pi}(\Delta')\times\nonumber\\
&&\sum_{\Delta''\in
T(\epsilon)}\epsilon^{imn}\mathrm{Tr}\big(A(s_i(\Delta''))^{-1}
\{A(s_i(\Delta'')),\sqrt{V_{U^\epsilon_{v(\Delta'')}}}\}\times\nonumber\\
&&A(s_m(\Delta''))^{-1}\{A(s_m(\Delta'')),\sqrt{V_{U^\epsilon_{v(\Delta'')}}}\}\times\nonumber\\
&&A(s_n(\Delta''))^{-1}\{A(s_n(\Delta'')),\sqrt{V_{U^\epsilon_{v(\Delta'')}}}\}\big)\times\nonumber\\
&&\sum_{\Delta'''\in
T(\epsilon)}\epsilon^{jkl}\mathrm{Tr}\big(A(s_j(\Delta'''))^{-1}
\{A(s_j(\Delta''')),\sqrt{V_{U^\epsilon_{v(\Delta''')}}}\}\times\nonumber\\
&&A(s_k(\Delta'''))^{-1}\{A(s_k(\Delta''')),\sqrt{V_{U^\epsilon_{v(\Delta''')}}}\}\times\nonumber\\
&&A(s_l(\Delta'''))^{-1}\{A(s_l(\Delta''')),\sqrt{V_{U^\epsilon_{v(\Delta''')}}}\}\big)\times\nonumber\\
&&\chi_\epsilon(v(\Delta)-v(\Delta'))\chi_\epsilon(v(\Delta'')-v(\Delta))\chi_\epsilon(v(\Delta''')-v(\Delta')),\label{classicalH}
\end{eqnarray}
where the overall numerical factors are got from the scale of a
tetrahedron to an octahedron. Note that the above regularization
method is in complete analogy with that used for Higgs field in
Ref.\cite{thiemann7}. However, our regularization formula of scalar
field Hamiltonian is explicitly dependent on the parameter
$\lambda$, which will leads to a kind of quantization ambiguity of
the real scalar field dynamics in polymer-like representation.

\subsection{Quantization of the Hamiltonian}

Since all constituents in the expression (\ref{classicalH}) have
clear quantum analogs, one can quantize it as an operator by
replacing the constituents by the corresponding operators and
Poisson brackets by canonical commutators. Then the regulator should
be removed by $\epsilon\rightarrow0$ with respect to a suitable
operator topology. Now we begin to construct the Hamiltonian
operator. Given a spin-scalar-network function $\Pi_{s,c}$, in order
to ensure that the final operator is diffeomorphism covariant, and
cylindrically consistent (up to diffeomorphisms), one can make the
triangulation $T(\epsilon)$ adapted to the graph $\gamma(s,c)$ of
$\Pi_{s,c}$ according to the strategy developed in
Ref.\cite{thiemann1} with the following properties.
\begin{itemize}
\item The graph $\gamma(s,c)$ is embedded in $T(\epsilon)$ for all
$\epsilon$, so that every vertex $v$ of $\gamma(s,c)$ coincides with
a vertex $v(\Delta)$ in $T(\epsilon)$.

\item For every triple of edges ($e_1,\ e_2,\ e_3$) of $\gamma(s,c)$
such that $v=s(e_1)=s(e_2)=s(e_3)$, there is a tetrahedra $\Delta\in
T(\epsilon)$ such that $v=v(\Delta)$ and $s_i(\Delta)\subset e_i,\
\forall\ i=1,2,3$. We denote such a tetrahedra as
$\Delta^0_{s_1,s_2,s_3}$.

\item For each tetrahedra $\Delta^0_{s_1,s_2,s_3}$ one can
construct seven additional tetrahedron
$\Delta^\wp_{s^\wp_1,s^\wp_2,s^\wp_3},\ \wp=1,...,7$, by backward
analytic extensions of $s_i(\Delta)$ so that
$U_{s_1,s_2,s_3}:=\cup_{\wp=0}^7\Delta^\wp_{s^\wp_1,s^\wp_2,s^\wp_3}$
is a neighborhood of $v$.

\item The triangulation must be fine enough so that the
neighborhoods $U(v):=\cup_{s_1,s_2,s_3}U_{s_1,s_2,s_3}(v)$ are
disjoint for different vertices $v$ and $v'$ of $\gamma(s,c)$. Thus
for any open neighborhood $U_{\gamma(s,c)}$ of the graph
$\gamma(s,c)$, there exists a triangulation $T(\epsilon)$ such that
$\cup_{v\in V(\gamma(s,c))}U(v)\subseteq U_{\gamma(s,c)}$.

\item The distance between a vertex $v(\Delta)$ and the
corresponding arcs $a_{ij}(\Delta)$ is described by the parameter
$\epsilon$. For any two different $\epsilon$ and $\epsilon'$, the
arcs $a_{ij}(\Delta^\epsilon)$ and $a_{ij}(\Delta^{\epsilon'})$ with
respect to one vertex $v(\Delta)$ are analytically diffeomorphic
with each other.

\item With the triangulation $T(\epsilon)$, the integral over
$\Sigma$ is replaced by the Riemanian sum:
\begin{eqnarray}
\int_\Sigma\ \ \ &=&\int_{U_{\gamma(s,c)}}\ \ \ \ +\int_{\Sigma-U_{\gamma(s,c)}}\ ,\ \ \ \nonumber\\
\int_{U_{\gamma(s,c)}}\ \ \ &=&\sum_{v\in V(\gamma(s,c))}\int_{U(v)}\ \ \ \ +\int_{U_{\gamma(s,c)}-\cup_{v}U(v)}\ ,\ \ \ \nonumber\\
\int_{U(v)}\ \ \
&=&\frac{1}{C^3_{n(v)}}\sum_{s_1,s_2,s_3}[\int_{U_{s_1,s_2,s_3}(v)}\
\ \ \ +\int_{U(v)-U_{s_1,s_2,s_3},(v)}\ ],\nonumber
\end{eqnarray}
where $n(v)$ is the valence of the vertex $v=s(s_1)=s(s_2)=s(s_3)$,
and $C^3_{n}\equiv \left(
\begin{array}{ll}
n\\
3
\end{array} \right)$ denotes the binomial coefficient. One then
observes that
\begin{eqnarray}
\int_{U_{s_1,s_2,s_3}(v)}\ \ \ \ =8\int_{\Delta^0_{s_1,s_2,s_3}(v)}\
\ \,\nonumber
\end{eqnarray}
in the limit $\epsilon\rightarrow0$.

\item The triangulation for the regions
\begin{eqnarray}
&&U(v)-U_{s_1,s_2,s_3}(v),\nonumber\\
&&U_{\gamma(s,c)}-\cup_{v\in V(\gamma(s,c))}U(v),\nonumber\\
&&\Sigma-U_{\gamma(s,c)}, \label{*}
\end{eqnarray}
are arbitrary. These regions do not contribute to the construction
of the operator, since the commutator terms like
$[\hat{A}(s_i(\Delta)),\hat{V}_{U_{v(\Delta)}}]\Pi_{s,c}$ would
vanish for all tetrahedron $\Delta$ in the regions (\ref{*}).
\end{itemize}
Note that there are many possible ways in choosing a triangulation.
The motivation for above choice has been fully discussed in
Ref.\cite{thiemann1}. Introducing a partition $\mathcal{P}$ of the
3-manifold $\Sigma$ into cells $C$, we can smear the "square roots"
of $\mathcal{H}^\epsilon_{KG,\phi}$ and
$\mathcal{H}^\epsilon_{KG,Kin}$ in one cell $C$ respectively and
promote them as regulated operators in $\mathcal{H}_{kin}$ with
respect to the state-dependent triangulation $T(\epsilon)$ as
\begin{eqnarray}
\hat{W}^{\epsilon,C}_{\gamma(s,c),\phi,i}&=&\sum_{v\in
V(\gamma(s,c))}\chi_C(v)\sum_{v(\Delta)=v}\frac{\hat{p}_{\Delta}}{\sqrt{\hat{E}(v)}}\hat{h}^{\epsilon,\Delta}_{\phi,v,i}
\frac{\hat{p}_{\Delta}}{\sqrt{\hat{E}(v)}},\nonumber\\
\hat{W}^{\epsilon,C}_{\gamma(s,c),Kin}&=&\sum_{v\in
V(\gamma(s,c))}\chi_C(v)\sum_{v(\Delta)=v}\frac{\hat{p}_{\Delta}}{\sqrt{\hat{E}(v)}}\hat{h}^{\epsilon,\Delta}_{Kin,v}
\frac{\hat{p}_{\Delta}}{\sqrt{\hat{E}(v)}}, \label{sqarerootH}
\end{eqnarray}
where $\chi_C(v)$ is the characteristic function of the cell $C$,
and
\begin{eqnarray}
\hat{h}^{\epsilon,\Delta}_{\phi,v,i}&:=&\frac{i}{\hbar^2}\epsilon^{lmn}\frac{1}{\lambda(v)}\hat{U}(v,\lambda(v))^{-1}[\hat{U}(t(s_l(\Delta)),\lambda(v))
-\hat{U}(v,\lambda(v))]\times\nonumber\\
&&\mathrm{Tr}\big(\tau_i\hat{A}(s_m(\Delta))^{-1}[\hat{A}(s_m(\Delta)),\hat{V}_{U^\epsilon_{v}}^{3/4}]
\hat{A}(s_n(\Delta))^{-1}[\hat{A}(s_n(\Delta)),\hat{V}_{U^\epsilon_{v}}^{3/4}]\big),\nonumber\\
\hat{h}^{\epsilon,\Delta}_{Kin,v}&:=&\frac{1}{(i\hbar)^3}\hat{\pi}(v)\epsilon^{lmn}\mathrm{Tr}\big(\hat{A}(s_l(\Delta))^{-1}[\hat{A}(s_l(\Delta)),
\sqrt{\hat{V}_{U^\epsilon_{v}}}]\times\nonumber\\
&&\hat{A}(s_m(\Delta))^{-1}[\hat{A}(s_m(\Delta)),\sqrt{\hat{V}_{U^\epsilon_{v}}}]\times\nonumber\\
&&\hat{A}(s_n(\Delta))^{-1}[\hat{A}(s_n(\Delta)),\sqrt{\hat{V}_{U^\epsilon_{v}}}]\big).\label{ambiguity}
\end{eqnarray}
Note that the tetrahedron projector $\hat{p}_{\Delta}$ associated
with segments $s_1$, $s_2$ and $s_3$ reads
\begin{eqnarray}
\hat{p}_{\Delta}&:=&\hat{p}_{s_1}\hat{p}_{s_2}\hat{p}_{s_3}\nonumber\\
&=&\theta(\sqrt{\frac{1}{4}-\Delta_{s_1}}-\frac{1}{2})
\theta(\sqrt{\frac{1}{4}-\Delta_{s_2}}-\frac{1}{2})\theta(\sqrt{\frac{1}{4}-\Delta_{s_3}}-\frac{1}{2})
\end{eqnarray}
where $\Delta_{s_i}$ is the Casimir operator associated with the
segment $s_i$ and $\theta$ is the distribution on $\mathbf{R}$ which
vanishes on $(-\infty,0]$ and equals $1$ on $(0,\infty)$, and the
tetrahedron projector related to a same vertex constitute the vertex
operator $\hat{E}(v):=\sum_{v(\Delta)=v}\hat{p}_{\Delta}$. Note also
that the partition $\mathcal{P}$ is not required to coincide with
the triangulation $T(\epsilon)$. We have arranged the operator
${\hat{p}_{\Delta}}/{\sqrt{\hat{E}(v)}}$ in such a way that both
operators in (\ref{sqarerootH}) and their adjoint operators are
cylindrically consistent up to diffeomorphisms. Thus there are two
densely defined operators $\hat{W}^{C}_{\phi,i}$ and
$\hat{W}^{C}_{Kin}$ in $\mathcal{H}_{kin}$ associated with the two
consistent families of (\ref{sqarerootH}). We now give several
remarks on their properties.
\begin{itemize}
\item {\textit{Removal of regulator $\epsilon$}}

It is not difficult to see that the action of the operator
$\hat{W}^{\epsilon,C}_{\gamma(s,c),\phi,i}$ on a
spin-scalar-network function $\Pi_{s,c}$ is graph-changing. It
adds finite number of vertices with representation $\lambda(v)$ at
$t(s_i(\Delta))$ with distance $\epsilon$ from the vertex $v$.
Recall that the action of the gravitational Hamiltonian constraint
operator on a spin network function is also graph-changing. As a
result, the family of operators
$\hat{W}^{\epsilon,C}_{\gamma(s,c),\phi,i}$ also fails to be
weakly convergent when $\epsilon\rightarrow0$. However, due to the
diffeomorphism covariant properties of the triangulation, the
limit operator can be well defined via the uniform Rovelli-Smolin
topology, or equivalently, the operator can be dually defined on
diffeomorphism invariant states. But the dual operator cannot
leave $\mathcal{H}_{Diff}$ invariant.

\item {\textit{Quantization ambiguity}}

As a main difference of the dynamics in polymer-like representation
from that in U(1) group representation \cite{thiemann7}, a
continuous label $\lambda$ appears explicitly in the expression of
(\ref{sqarerootH}). Hence there is an one-parameter quantization
ambiguity due to the real scalar field. Recall that the construction
of gravitational Hamiltonian constraint operator also has a similar
ambiguity due to the choice of the representations $j$ of the edges
added by its action. A related quantization ambiguity also appears
in the dynamics of loop quantum cosmology \cite{boj}. We will come
back to this point in a future publication \cite{kernel}. Since the
regulator is removed in a diffeomorphism invariant way, the
quadratic form, which we are going to constructed, will be
independent of the initial triangulation $T$ in the sense that it
depends only on the diffeomorphism class of $T$, as is the case of
the gravitational Hamiltonian constraint operator \cite{thiemann1}.
\end{itemize}
Since our quantum field theory is expected to be diffeomorphism
invariant, we would like to define the Hamiltonian operator of
polymer scalar field in the diffeomorphism invariant Hilbert space
$\mathcal{H}_{Diff}$. For this purpose we fix the parameter
$\lambda$ to be a non-zero constant at every point. Then what we
will do is to employ the new quantization strategy developed in
Refs. \cite{thiemann3} and \cite{thiemann15}. We first construct a
quadratic form in the light of a new inner product defined in
Ref.\cite{thiemann15} on the algebraic dual $\mathcal{D}^\star$ of
the space of cylindrical functions. Then we prove that the quadratic
form is closed. Note that, although the calculation by employing
this inner product is formal, it can led to a well-defined
expression of the desired quadratic form Eq.(\ref{quadrticform4}).
Since an arbitrary element of $\mathcal{D}^\star$ is of the form
$\Psi=\sum_{s,c} c_{s,c}<\Pi_{s,c}|\ \cdot>_{kin}$, one can formally
define an inner product $<\cdot\ |
 \cdot>_\star$ on
$\mathcal{D}^\star$ via
\begin{eqnarray}
<\Psi,\Psi'>_\star&:=&<\sum_{s,c} c_{s,c}<\Pi_{s,c}|\
\cdot>_{kin}|\sum_{s',c'}
c'_{s',c'}<\Pi_{s',c'}|\ \cdot>_{kin}>_\star\nonumber\\
&:=&\sum_{s,c;s',c'}c_{s,c}\overline{c'_{s',c'}}<\Pi_{s,c}|\Pi_{s',c'}>_{kin}
\frac{1}{\sqrt{\aleph([s,c])\aleph([s',c'])}}\nonumber\\
&=&\sum_{s,c}c_{s,c}\overline{c'_{s,c}}\frac{1}{\aleph([s,c])},\label{product}
\end{eqnarray}
where the Cantor aleph $\aleph$ denotes the cardinal of the set
$[s,c]$. Note that we exchange the coefficients on which the complex
conjugate was taken in Ref.\cite{thiemann15}, so that the inner
product $<\Psi_{Diff} |\Psi'_{Diff}>_\star$ reduces to
$<\Psi_{Diff}|\Psi'_{Diff}>_{Diff}$ for any
$\Psi_{Diff},\Psi'_{Diff}\in\mathcal{H}_{Diff}$. Completing the
quotient with respect to the null vectors by this inner product, one
gets a Hilbert space $\mathcal{H}_\star$. Our purpose is to
construct a quadratic form associated to some positive and symmetric
operator in analogy with the classical expression of
(\ref{classicalH}). So the quadratic form should first be given in a
positive and symmetric version. It is then natural to define two
quadratic forms on a dense subset of
$\mathcal{H}_{Diff}\subset\mathcal{H}_\star$ as:
\begin{eqnarray}
Q_{KG,\phi}(\Psi_{Diff},
\Psi'_{Diff})&:=&\lim_{\mathcal{P}\rightarrow\Sigma}\sum_{C\in\mathcal{P}}64\times
\frac{4\alpha_M}{9\kappa^4\gamma^4}\delta^{ij}<\hat{W}'^{C}_{\phi,i}\Psi_{Diff}|\hat{W}'^{C}_{\phi,j}\Psi'_{Diff}>_\star,\nonumber\\
Q_{KG,Kin}(\Psi_{Diff},
\Psi'_{Diff})&:=&\lim_{\mathcal{P}\rightarrow\Sigma}\sum_{C\in\mathcal{P}}8^4\times
\frac{16}{81\alpha_M(\kappa\gamma)^6}<\hat{W}'^{C}_{Kin}\Psi_{Diff}|\hat{W}'^{C}_{Kin}\Psi'_{Diff}>_\star,\nonumber\\
\label{quadrticform1}
\end{eqnarray}
where the dual limit operator $\hat{W}'^{C}$ of either family of
$\hat{W}^{\epsilon,C}_{\phi,i}$ or $\hat{W}^{\epsilon,C}_{Kin}$ in
(\ref{sqarerootH}) is naturally defined on diffeomorphism invariant
states as
\begin{eqnarray}
\hat{W}'^{C}\Psi_{Diff}[\Pi_{s,c}]=\lim_{\epsilon\rightarrow0}\Psi_{Diff}[\hat{W}^{\epsilon,C}\Pi_{s,c}].
\end{eqnarray}
To show that the quadratic forms are well defined, we write
\begin{eqnarray}
\hat{W}'^{C}_{\phi,i}\Psi_{Diff}=\sum_{s,c}w^{\Psi}_{\phi,i,s,c}(C)<\Pi_{s,c}|\
\cdot>_\star&\Rightarrow&
w^{\Psi}_{\phi,i,s,c}(C)=(\hat{W}'^{C}_{\phi,i}\Psi_{Diff})[\Pi_{s,c}],\nonumber\\
\hat{W}'^{C}_{Kin}\Psi_{Diff}=\sum_{s,c}w^{\Psi}_{Kin,s,c}(C)<\Pi_{s,c}|\
\cdot>_\star&\Rightarrow&
w^{\Psi}_{Kin,s,c}(C)=(\hat{W}'^{C}_{Kin}\Psi_{Diff})[\Pi_{s,c}].\nonumber
\end{eqnarray}
Then, by using the inner product (\ref{product}) the quadratic forms
in (\ref{quadrticform1}) become
\begin{eqnarray}
&&Q_{KG,\phi}(\Psi_{Diff},\Psi'_{Diff})\nonumber\\
&:=&\lim_{\mathcal{P}\rightarrow\Sigma}\sum_{C\in\mathcal{P}}64\times
\frac{4\alpha_M}{9\kappa^4\gamma^4}\delta^{ij}\sum_{s,c}
w^{\Psi}_{\phi,i,s,c}(C)\overline{w^{\Psi'}_{\phi,j,s,c}(C)}\frac{1}{\aleph([s,c])}\nonumber\\
&=&\lim_{\mathcal{P}\rightarrow\Sigma}\sum_{C\in\mathcal{P}}64\times
\frac{4\alpha_M}{9\kappa^4\gamma^4}\delta^{ij}\sum_{[s,c]}\frac{1}{\aleph([s,c])}\sum_{s,c\in[s,c]}
w^{\Psi}_{\phi,i,s,c}(C)\overline{w^{\Psi'}_{\phi,j,s,c}(C)},\nonumber\\
&&\overline{Q_{KG,Kin}(\Psi_{Diff},\Psi'_{Diff})}\nonumber\\
&:=&\lim_{\mathcal{P}\rightarrow\Sigma}\sum_{C\in\mathcal{P}}8^4\times
\frac{16}{81\alpha_M(\kappa\gamma)^6}\sum_{s,c}
w^{\Psi}_{Kin,s,c}(C)\overline{w^{\Psi'}_{Kin,s,c}(C)}\frac{1}{\aleph([s,c])}\nonumber\\
&=&\lim_{\mathcal{P}\rightarrow\Sigma}\sum_{C\in\mathcal{P}}8^4\times
\frac{16}{81\alpha_M(\kappa\gamma)^6}\sum_{[s,c]}\frac{1}{\aleph([s,c])}\sum_{s,c\in[s,c]}
w^{\Psi}_{Kin,s,c}(C)\overline{w^{\Psi'}_{Kin,s,c}(C)}.\nonumber\\
\label{quadrticform2}
\end{eqnarray}
Note that, since $\Psi_{Diff}$ is a finite linear combination of the
diffeomorphism invariant spin-scalar-network basis, taking account
of the operational property of $\hat{W}'^{C}$ there are only finite
number of terms in the summation $\sum_{[s,c]}$ contributing to
(\ref{quadrticform2}). Hence we can interchange $\sum_{[s,c]}$ and
$\lim_{\mathcal{P}\rightarrow\Sigma}\sum_{C\in\mathcal{P}}$ in above
calculation. Moreover, for a sufficiently fine partition such that
each cell contains at most one vertex, the sum over cells therefore
reduces to finite terms with respect to the vertices of
$\gamma(s,c)$. So we can interchange $\sum_{s,c\in[s,c]}$ and
$\lim_{\mathcal{P}\rightarrow\Sigma}\sum_{C\in\mathcal{P}}$ to
obtain:
\begin{eqnarray}
&&Q_{KG,\phi}(\Psi_{Diff},\Psi'_{Diff})\nonumber\\
&=&64\times\frac{4\alpha_M}{9\kappa^4\gamma^4}\delta^{ij}
\sum_{[s,c]}\frac{1}{\aleph([s,c])}\sum_{s,c\in[s,c]}
\lim_{\mathcal{P}\rightarrow\Sigma}\sum_{C\in\mathcal{P}}
w^{\Psi}_{\phi,i,s,c}(C)\overline{w^{\Psi'}_{\phi,j,s,c}(C)}\nonumber\\
&=&64\times\frac{4\alpha_M}{9\kappa^4\gamma^4}\delta^{ij}
\sum_{[s,c]}\frac{1}{\aleph([s,c])}\sum_{s,c\in[s,c]} \sum_{v\in
V(\gamma(s,c))}
(\hat{W}'^{v}_{\phi,i}\Psi_{Diff})[\Pi_{s,c}]\overline{(\hat{W}'^{v}_{\phi,j}\Psi'_{Diff})[\Pi_{s,c}]},\nonumber\\
&&Q_{KG,Kin}(\Psi_{Diff},\Psi'_{Diff})\nonumber\\
&=&8^4\times\frac{16}{81\alpha_M(\kappa\gamma)^6}
\sum_{[s,c]}\frac{1}{\aleph([s,c])}\sum_{s,c\in[s,c]}
\lim_{\mathcal{P}\rightarrow\Sigma}\sum_{C\in\mathcal{P}}
w^{\Psi}_{Kin,s,c}(C)\overline{w^{\Psi'}_{Kin,s,c}(C)}\nonumber\\
&=&8^4\times\frac{16}{81\alpha_M(\kappa\gamma)^6}
\sum_{[s,c]}\frac{1}{\aleph([s,c])}\sum_{s,c\in[s,c]} \sum_{v\in
V(\gamma(s,c))}(\hat{W}'^{v}_{Kin}\Psi_{Diff})[\Pi_{s,c}]\overline{(\hat{W}'^{v}_{Kin}\Psi'_{Diff})[\Pi_{s,c}]},\nonumber\\
\label{quadrticform3}
\end{eqnarray}
where the limit $\mathcal{P}\rightarrow\Sigma$ has been taken so
that $C\rightarrow v$. Since given $\gamma(s,c)$ and $\gamma(s',c')$
which are different up to a diffeomorphism transformation, there is
always a diffeomorphism $\varphi$ transforming the graph associated
with $\hat{W}^{\epsilon,v}_{\gamma(s,c)}\Pi_{s,c}\
(v\in\gamma(s,c))$ to that of
$\hat{W}^{\epsilon,v'}_{\gamma(s',c')}\Pi_{s',c'}\
(v'\in\gamma(s',c'))$ with $\varphi(v)=v'$,
$(\hat{W}'^{v}\Psi_{Diff})[\Pi_{s,c\in[s,c]}]$ is constant for
different $(s,c)\in[s,c]$, i.e., all the $\aleph([s,c])$ terms in
the sum over $(s,c)\in[s,c]$ are identical. Hence the final
expressions of the two quadratic forms can be written as:
\begin{eqnarray}
&&Q_{KG,\phi}(\Psi_{Diff},\Psi'_{Diff})\nonumber\\
&=&64\times\frac{4\alpha_M}{9\kappa^4\gamma^4}\delta^{ij}
\sum_{[s,c]}\sum_{v\in V(\gamma(s,c))}
(\hat{W}'^{v}_{\phi,i}\Psi_{Diff})[\Pi_{s,c\in[s,c]}]\overline{(\hat{W}'^{v}_{\phi,j}\Psi'_{Diff})[\Pi_{s,c\in[s,c]}]},\nonumber\\
&&Q_{KG,Kin}(\Psi_{Diff},\Psi'_{Diff})\nonumber\\
&=&8^4\times\frac{16}{81\alpha_M(\kappa\gamma)^6}
\sum_{[s,c]}\sum_{v\in V(\gamma(s,c))}
(\hat{W}'^{v}_{Kin}\Psi_{Diff})[\Pi_{s,c\in[s,c]}]\overline{(\hat{W}'^{v}_{Kin}\Psi'_{Diff})[\Pi_{s,c\in[s,c]}]}.\nonumber\\
\label{quadrticform4}
\end{eqnarray}
Note that both quadratic forms in (\ref{quadrticform4}) have finite
results and hence their form domains are dense in
$\mathcal{H}_{Diff}$. Moreover, both of them are obviously positive, and the following theorem will demonstrate their closedness. \\ \\
\textbf{Theorem}: \textit{Both $Q_{KG,\phi}$ and $Q_{KG,Kin}$ are
densely defined, positive and closed quadratic forms on
$\mathcal{H}_{Diff}$, which are associated uniquely with two
positive self-adjoint operators  respectively on
$\mathcal{H}_{Diff}$ such that
\begin{eqnarray}
Q_{KG,\phi}(\Psi_{Diff},\Psi'_{Diff})&=&<\Psi_{Diff}|\hat{\mathcal{H}}_{KG,\phi}|\Psi'_{Diff}>_{Diff}\nonumber\\
Q_{KG,Kin}(\Psi_{Diff},\Psi'_{Diff})&=&<\Psi_{Diff}|\hat{\mathcal{H}}_{KG,Kin}|\Psi'_{Diff}>_{Diff}.\nonumber
\end{eqnarray}
Therefore the Hamiltonian operator
\begin{eqnarray}
\hat{\mathcal{H}}_{KG}:=\hat{\mathcal{H}}_{KG,\phi}+\hat{\mathcal{H}}_{KG,Kin}\label{HKG}
\end{eqnarray}
is positive and also have a unique self-adjoint extension.}\\ \\
\textbf{Proof}: We follow the strategy developed in
Refs.\cite{thiemann15} and \cite{HM2} to prove that both
$Q_{KG,\phi}$ and $Q_{KG,Kin}$ are closeable and uniquely induce two
positive self-adjoint operators $\hat{\mathcal{H}}_{KG,\phi}$ and
$\hat{\mathcal{H}}_{KG,Kin}$. One can formally define
$\hat{\mathcal{H}}_{KG,\phi}$ and $\hat{\mathcal{H}}_{KG,Kin}$
acting on diffeomorphism invariant spin-scalar network functions
via:
\begin{eqnarray}
\hat{\mathcal{H}}_{KG,\phi}\ \Pi_{[s_1,c_1]}&:=&\sum_{[s_2,c_2]}Q_{KG,\phi}(\Pi_{[s_2,c_2]},\Pi_{[s_1,c_1]})\Pi_{[s_2,c_2]},\label{define1}\\
\hat{\mathcal{H}}_{KG,Kin}\
\Pi_{[s_1,c_1]}&:=&\sum_{[s_2,c_2]}Q_{KG,Kin}(\Pi_{[s_2,c_2]},\Pi_{[s_1,c_1]})\Pi_{[s_2,c_2]}.\label{define2}
\end{eqnarray}
Then we need to show that both of the above operators are densely
defined on the Hilbert space $\mathcal{H}_{Diff}$, i.e.,
\begin{eqnarray}
||\hat{\mathcal{H}}_{KG,\phi}\Pi_{[s_1,c_1]}||_{Diff}&=&\sum_{[s_2,c_2]}|Q_{KG,\phi}(\Pi_{[s_2,c_2]},\Pi_{[s_1,c_1]})|^2<\infty,\label{dense1}\\
||\hat{\mathcal{H}}_{KG,Kin}\Pi_{[s_1,c_1]}||_{Diff}&=&\sum_{[s_2,c_2]}|Q_{KG,Kin}(\Pi_{[s_2,c_2]},\Pi_{[s_1,c_1]})|^2<\infty.\label{dense2}
\end{eqnarray}
Given a diffeomorphism invariant spin-scalar network function
$\Pi_{[s_1,c_1]}$, there are only finite number of terms
$\Pi_{[s_1,c_1]}
[\hat{W}^{\epsilon,v}_{\gamma(s,c)}\Pi_{s,c\in[s,c]}]$ which are
nonzero in the sum over equivalent classes $[s,c]$ in
(\ref{quadrticform4}). On the other hand, given one spin-scalar
network function $\Pi_{s,c\in[s,c]}$, there are also only finite
number of possible $\Pi_{[s_2,c_2]}$ such that the terms
$\overline{\Pi_{[s_2,c_2]}[\hat{W}^{\epsilon,v}_{\gamma(s,c)}\Pi_{s,c\in[s,c]}]}$
are nonzero. As a result, only finite number of terms survive in
both sums over $[s_2,c_2]$ in Eqs. (\ref{dense1}) and
(\ref{dense2}). Hence both $\hat{\mathcal{H}}_{KG,\phi}$ and
$\hat{\mathcal{H}}_{KG,Kin}$ are well defined. Then it follows from
Eqs. (\ref{quadrticform4}), (\ref{define1}) and (\ref{define2}) that
they are positive and symmetric operators densely defined in
$\mathcal{H}_{Diff}$, whose quadratic forms coincide with
$Q_{KG,\phi}$ and $Q_{KG,Kin}$ on their form domains. Hence both
$Q_{KG,\phi}$ and $Q_{KG,Kin}$ have positive closures and uniquely
induce self-adjoint (Friedrichs) extensions of
$\hat{\mathcal{H}}_{KG,\phi}$ and $\hat{\mathcal{H}}_{KG,Kin}$
respectively \cite{rs}, which we denote by
$\hat{\mathcal{H}}_{KG,\phi}$ and $\hat{\mathcal{H}}_{KG,Kin}$ as
well. As a result, the Hamiltonian operator $\hat{\mathcal{H}}_{KG}$
defined by Eq.(\ref{HKG}) is also positive and symmetric. Hence it
has a unique
self-adjoint (Friedrichs) extension.\\
$\Box$

We notice that, from a different perspective, one can construct the
same Hamiltonian operator $\hat{\mathcal{H}}_{KG}$ without
introducing an inner product on $\mathcal{D}^\star$. The
construction is sketched as follows. Using the two well-defined
operators $\hat{W}^{\epsilon,C}_{\phi,i}$ and
$\hat{W}^{\epsilon,C}_{Kin}$ as in (\ref{sqarerootH}), as well as
their adjoint operators $(\hat{W}^{\epsilon,C}_{\phi,i})^\dagger$
and $(\hat{W}^{\epsilon,C}_{Kin})^\dagger$, one may define two
operators on $\mathcal{H}_{Diff}$ corresponding to the two terms in
(\ref{classicalH}) by
\begin{eqnarray}
(\hat{\mathcal{H}}_{KG,\phi}\Psi_{Diff})[f_\gamma]&=&\lim_{\epsilon,\epsilon'\rightarrow0,\mathcal{P}\rightarrow\Sigma}
\Psi_{Diff}[\sum_{C\in\mathcal{P}}64\times
\frac{4\alpha_M}{9\kappa^4\gamma^4}\delta^{ij}\hat{W}^{\epsilon,C}_{\phi,i}(\hat{W}^{\epsilon',C}_{\phi,j})^\dagger
f_\gamma]\nonumber\\
(\hat{\mathcal{H}}_{KG,Kin}\Psi_{Diff})[f_\gamma]&=&\lim_{\epsilon,\epsilon'\rightarrow0,\mathcal{P}\rightarrow\Sigma}
\Psi_{Diff}[\sum_{C\in\mathcal{P}}8^4\times
\frac{16}{81\alpha_M(\kappa\gamma)^6}\hat{W}^{\epsilon,C}_{Kin}(\hat{W}^{\epsilon',C}_{Kin})^\dagger
f_\gamma].\nonumber\\
\end{eqnarray}
In analogy with the discussion in section 5 and Ref.\cite{HM2}, it
can be shown that both above operators leave $\mathcal{H}_{Diff}$
invariant and are densely defined on $\mathcal{H}_{Diff}$. Moreover,
the quadratic forms associated with them coincide with the quadratic
forms in (\ref{quadrticform4}). Thus the Hamiltonian operator
$\hat{\mathcal{H}}_{KG}:=\hat{\mathcal{H}}_{KG,\phi}+\hat{\mathcal{H}}_{KG,Kin}$
coincides with the one constructed in the quadratic form approach.

In summary, we have constructed a positive self-adjoint Hamiltonian
operator on $\mathcal{H}_{Diff}$ for the polymer-like scalar field,
depending on a chosen parameter $\lambda$. Thus there is an
1-parameter ambiguity in the construction. However, there is no UV
divergence in this quantum Hamiltonian without renormalization,
since quantum gravity is presented as a natural regulator for the
polymer-like scalar field.

\section{Quantum Hamiltonian constraint equation}

In this section we consider the whole dynamical system of scalar
field coupled to gravity. Recall that in perturbative quantum
field theory in curved spacetime, the definition of some basic
physical quantities, such as the expectation value of
energy-momentum, is ambiguous and it is extremely difficult to
calculate the back-reaction of quantum fields to the background
spacetime \cite{wald}. This is reflected by the fact that the
semi-classical Einstein equation,
\begin{equation}
R_{\alpha\beta}[g]-\frac{1}{2}R[g]g_{\alpha\beta}=\kappa
<\hat{T}_{\alpha\beta}[g]>,\label{ein}
\end{equation}
is inconsistent and ambiguous \cite{FM}\cite{thiemann2}. One could
speculate on that the difficulty is related to the fact that the
present formulation of quantum field theories are background
dependent. According to this speculation, if the quantization
programme is by construction non-perturbative and background
independent, it is possible to solve the problems fundamentally.
In loop quantum gravity, there is no assumption of a priori
background metric at all. The quantum geometry and quantum matter
fields are coupled and fluctuating naturally with respect to each
other on a common manifold. On the other hand, there exists the
"time problem" in quantum theory of pure gravity, since all the
physical states have to satisfy certain version of quantum
Wheeler-DeWitt constraint equation. However, the situation would
be improved when matter field is coupled to gravity. In the
following construction, we impose the quantum Hamiltonian
constraint on $\mathcal{H}_{kin}$, and thus define a quantum
Wheeler-DeWitt constraint equation for the scalar field coupled to
gravity. Then one can gain an insight into the problem of time
from the coupled equation, and the back-reaction of the quantum
scalar field is included in the framework of loop quantum gravity.

Recall that the gravitational Hamiltonian constraint operator
$\hat{\mathcal{H}}_{GR}(N)$ can be well defined in
$\mathcal{H}_{kin}^{GR}$ by the uniform Rovelli-Smolin topology
\cite{thiemann1}\cite{HHM}. Hence it is also well defined in the
coupled kinematical Hilbert space $\mathcal{H}_{kin}$. Its regulated
version via a state-dependent triangulation $T(\epsilon)$ reads
\begin{eqnarray}
\hat{\mathcal{H}}_{GR}^\epsilon(N)&=&\hat{\mathcal{H}}^\epsilon_{E}(N)-2(1+\gamma^2)
\hat{\mathcal{T}}^\epsilon(N)\nonumber\\
\hat{\mathcal{H}}^\epsilon_{E,\alpha}(N)&=&\frac{16}{3i\hbar\kappa^2\gamma}\sum_{v\in
V(\alpha)}
N(v)\sum_{v(\Delta)=v}\epsilon^{ijk}\times\nonumber\\
&&\mathrm{Tr}\big(\hat{A}(\alpha_{ij}(\Delta))^{-1}\hat{A}(s_k(\Delta))^{-1}[\hat{A}(s_k(\Delta)),
\hat{V}_{U^\epsilon_{v}}]\big)\frac{\hat{p}_{\Delta}}{\hat{E}(v)},\nonumber\\
\hat{\mathcal{T}}_\alpha^\epsilon(N)&=&-\frac{4\sqrt{2}}{3i\hbar^3\kappa^4\gamma^3}\sum_{v\in
V(\alpha)}N(v)\sum_{v(\Delta)=v}\epsilon^{ijk}\times\nonumber\\
&&\mathrm{Tr}\big(\hat{A}(s_i(\Delta))^{-1}[\hat{A}(s_i(\Delta)),\hat{K}^\epsilon]\hat{A}(s_j(\Delta))^{-1}
[\hat{A}(s_j(\Delta)),\hat{K}^\epsilon]\times\nonumber\\
&&\hat{A}(s_k(\Delta))^{-1}[\hat{A}(s_k(\Delta)),\hat{V}_{U^\epsilon_{v}}]\big)\frac{\hat{p}_{\Delta}}{\hat{E}(v)}.
\label{GRhamilton}
\end{eqnarray}
We now define an operator in $\mathcal{H}_{kin}$ corresponding to
the scalar field part $\mathcal{H}_{KG}(N)$ of the total Hamiltonian
constraint functional, which can be read out from Eqs.
(\ref{hamilton}) and (\ref{kgconstraint}) as
\begin{eqnarray}
\mathcal{H}_{KG}(N)&=&\mathcal{H}_{KG,\phi}(N)+\mathcal{H}_{KG,Kin}(N),\nonumber
\end{eqnarray}
where
\begin{eqnarray}
\mathcal{H}_{KG,\phi}(N)&=&\frac{\kappa^2\gamma^2\alpha_M}{2}\int_\Sigma
d^3xN\frac{1}{\sqrt{|\det
q|}}\delta^{ij}\widetilde{P}^{a}_{i}\widetilde{P}^{b}_{j}(\partial_a\phi)\partial_b\phi,\nonumber\\
\mathcal{H}_{KG,Kin}(N)&=&\frac{1}{2\alpha_M}\int_\Sigma
d^3xN\frac{1}{\sqrt{|\det q|}}\widetilde{\pi}^2\nonumber.
\end{eqnarray}
In analogy with the regularization and quantization in the previous
section, the regulated version of quantum Hamiltonian constraint
$\hat{\mathcal{H}}^\epsilon_{KG}(N)$ of scalar field is expressed
via a state-dependent triangulation $T(\epsilon)$ as:
\begin{eqnarray}
\hat{\mathcal{H}}^{\epsilon}_{KG,\gamma}(N):=\sum_{v\in
V(\gamma)}N(v)
[\delta^{ij}(\hat{W}^{\epsilon,v}_{\gamma,\phi,i})^\dagger\hat{W}^{\epsilon,v}_{\gamma,\phi,j}
+(\hat{W}^{\epsilon,v}_{\gamma,Kin})^\dagger\hat{W}^{\epsilon,v}_{\gamma,Kin}],\label{scalarconstraint}
\end{eqnarray}
where the operators
\begin{eqnarray}
\hat{W}^{\epsilon,v}_{\gamma,\phi,i}:=\sum_{v(\Delta)=v}\frac{\hat{p}_{\Delta}}{\sqrt{\hat{E}(v)}}
\hat{h}^{\epsilon,\Delta}_{\phi,v,i}
\frac{\hat{p}_{\Delta}}{\sqrt{\hat{E}(v)}}&,&
\hat{W}^{\epsilon,v}_{\gamma,Kin}:=\sum_{v(\Delta)=v}\frac{\hat{p}_{\Delta}}{\sqrt{\hat{E}(v)}}
\hat{h}^{\epsilon,\Delta}_{Kin,v}
\frac{\hat{p}_{\Delta}}{\sqrt{\hat{E}(v)}},\nonumber\\
(\hat{W}^{\epsilon,v}_{\gamma,\phi,i})^\dagger:=\sum_{v(\Delta)=v}\frac{\hat{p}_{\Delta}}{\sqrt{\hat{E}(v)}}
(\hat{h}^{\epsilon,\Delta}_{\phi,v,i})^\dagger\frac{\hat{p}_{\Delta}}{\sqrt{\hat{E}(v)}}
&,&
(\hat{W}^{\epsilon,v}_{\gamma,Kin})^\dagger:=\sum_{v(\Delta)=v}\frac{\hat{p}_{\Delta}}{\sqrt{\hat{E}(v)}}
(\hat{h}^{\epsilon,\Delta}_{Kin,v})^\dagger\frac{\hat{p}_{\Delta}}{\sqrt{\hat{E}(v)}}\nonumber
\end{eqnarray}
are all cylindrically consistent up to diffeomorphisms. Hence the
family of Hamiltonian constraint operators (\ref{scalarconstraint})
is also cylindrically consistent up to diffeomorphisms, and the
regulator $\epsilon$ can be removed via the uniform Rovelli-Smollin
topology, or equivalently the limit operator dually acts on
diffeomorphism invariant states as
\begin{eqnarray}
(\hat{\mathcal{H}}'_{KG}(N)\Psi_{Diff})[f_\gamma]=
\lim_{\epsilon\rightarrow0}\Psi_{Diff}[\hat{\mathcal{H}}^{\epsilon}_{KG,\gamma}(N)f_\gamma],
\end{eqnarray}
for any $f_{\gamma}\in
Cyl_{\gamma(s,c)}(\overline{\mathcal{A}})\otimes
Cyl_{\gamma(s,c)}(\overline{\mathcal{U}})$. Similar to the dual of
$\hat{\mathcal{H}}_{GR}(N)$, the operator
$\hat{\mathcal{H}}'_{KG}(N)$ fails to commute with the dual of
finite diffeomorphism transformation operators, unless the smearing
function $N(x)$ is a constant function over $\Sigma$. Note that the
diffeomorphism invariant Hamiltonian operator
$\hat{\mathcal{H}}_{KG}$ defined in the previous section is actually
$\hat{\mathcal{H}}'_{KG}(1)$. From Eq.(\ref{scalarconstraint}), it
is not difficult to prove that for positive $N(x)$ the Hamiltonian
constraint operator $\hat{\mathcal{H}}_{KG}(N)$ of scalar field is
positive and symmetric in $\mathcal{H}_{kin}$ and hence has a unique
self-adjoint extension. It is pointed out in Ref.\cite{kuchar} that,
there can be problems associated with symmetric constraint operators
for systems where the constraints close with structure functions as
is the present case. However, not all the assumptions underlying
this conclusion are valid in the framework of loop quantum gravity.
For example, it is assumed in Ref.\cite{kuchar} that all the
classical canonical variables and constraints could be promoted as
well-defined operators in the kinematical Hilbert space. However, it
is well known that the classical connection and diffeomorphsim
constraint cannot be represented as well-defined operators in loop
quantum gravity. This issue related to the symmetric Hamiltonian
constraint operator was fully discussed in Ref.\cite{thiemann1}.

Our construction of $\hat{\mathcal{H}}_{KG}(N)$ is similar to that
of the Higgs field Hamiltonian constraint in Ref.\cite{thiemann7}.
However, like the case of $\hat{\mathcal{H}}_{KG}$, there is an
1-parameter ambiguity in our construction of
$\hat{\mathcal{H}}_{KG}(N)$ due to the real scalar field, which is
manifested as the continuous parameter $\lambda$ in the expression
of $\hat{h}^{\epsilon,\Delta}_{\phi,v,i}$ in (\ref{ambiguity}). Note
that now $\lambda$ is not required to be a constant, i.e., its value
can be changed from one point to another. This issue of ambiguity
will be discussed again in a future publication \cite{kernel}. Thus
the total Hamiltonian constraint operator of scalar field coupled to
gravity has been obtained as
\begin{eqnarray}
\hat{\mathcal{H}}(N)=\hat{\mathcal{H}}_{GR}(N)+\hat{\mathcal{H}}_{KG}(N).\label{Hconstraint}
\end{eqnarray}
Again, there is no UV divergence in this quantum Hamiltonian
constraint. Recall that, in standard quantum field theory the UV
divergence can only be cured by renormalization procedure, in which
one has to multiply the Hamiltonian by a suitable power of the
regulating parameter $\epsilon$ artificially. While, now $\epsilon$
has naturally disappeared from the expressions of (\ref{HKG}) and
(\ref{Hconstraint}). So renormalization is not needed for the
polymer-like scalar field coupled to gravity, since quantum gravity
has been presented as a natural regulator. Together with the result
in the previous section, this heightens our confidence that the
issue of divergence in quantum fields theory can be cured in the
framework of loop quantum gravity. The desired matter-coupled
Wheeler-DeWitt equation can be well imposed as:
\begin{eqnarray}
-\big(\hat{\mathcal{H}}_{KG}'(N)\Psi_{Diff}\big)[f_\gamma]=\big(\hat{\mathcal{H}}_{GR}'(N)
\Psi_{Diff}\big)[f_\gamma].\label{evo constr}
\end{eqnarray}
Note that the scalar field part $\hat{\mathcal{H}}_{KG}(N)$ acts
nontrivially on gravitational quantum states. This can be regarded
as the "back-reaction" of quantum matter field to the quantum
gravitational field. On the other hand, comparing Eq.(\ref{evo
constr}) with the well-known Sch\"{o}rdinger equation for a
particle,
\begin{eqnarray}
{i\hbar\frac{\partial}{\partial
t}}\psi(x,t)=H(\hat{x},\widehat{-i\hbar\frac{\partial}{\partial
x}})\psi(x,t),\nonumber
\end{eqnarray}
where $\psi(x,t)\in L^2(\mathbf{R},dx)$ and $t$ is a parameter
labelling time evolution, one may take the viewpoint that the
matter field constraint operator $\hat{\mathcal{H}}_{KG}'(N)$
plays the role of $i\hbar\frac{\partial}{\partial t}$. Then $\phi$
appears as the parameter labelling the evolution of the
gravitational field state. In the reverse viewpoint, gravitational
field would become the parameter labelling the evolution of the
quantum matter field.

\section{Master constraint programme}

In order to avoid possible quantum anomaly and find the physical
Hilbert space of quantum gravity, master constraint programme was
first introduced by Thiemann in \cite{thiemann3}. The central idea
is to construct an alternative classical constraint algebra, giving
the same constraint phase space, which is a Lie algebra (no
structure function) and where the subalgebra of diffeomorphism
constraints forms an ideal. Self-adjoint master constraint operators
for loop quantum gravity are then proposed in Refs.
\cite{thiemann15} and \cite{HM2}. The master constraint programme
can be generalized to matter fields coupled to gravity in a
straightforward way. We now take the massless real scalar field to
demonstrate the construction of a master constraint operator
according to the strategy in Ref.\cite{HM2}. By this approach one
not only avoids possible quantum anomaly which might appear in the
conventional canonical quantization method, but also might give a
qualitative description of the physical Hilbert space for the
coupled system. We introduce the master constraint for the scalar
field coupled to gravity as
\begin{eqnarray}
\textbf{M}:=\frac{1}{2}\int_\Sigma d^3x\frac{|{C}(x)|^2}{\sqrt{|\det
q(x)|}},\label{mconstraint}
\end{eqnarray}
where ${C}(x)$ is the Hamiltonian constraint in
(\ref{kgconstraint}). After solving the Gaussian constraint, one
gets the master constraint algebra as a Lie algebra:
\begin{eqnarray}
\{\mathcal{V}(\vec{N}),\ \mathcal{V}(\vec{N}')\}&=&\mathcal{V}([\vec{N},\vec{N}']),\nonumber\\
\{\mathcal{V}(\vec{N}),\ \textbf{M}\}&=&0,\nonumber\\
\{\textbf{M},\ \textbf{M}\}&=&0,\label{malgebra}
\end{eqnarray}
where the subalgebra of diffeomorphism constraints forms an ideal.
So it is possible to define a corresponding master constraint
operator on $\mathcal{H}_{Diff}$. In the following, the positivity
and the diffeomorphism invariance of $\textbf{M}$ will be working
together properly and provide us with powerful functional analytic
tools in the quantization procedure.

The regulated version of the master constraint can be expressed via
a point-splitting strategy as:
\begin{eqnarray}
\textbf{M}^{\epsilon}:=\frac{1}{2}\int_\Sigma d^3y \int_\Sigma
d^3x\chi_\epsilon(x-y)\frac{C(y)}{\sqrt{V_{U_y^\epsilon}}}
\frac{{C}(x)}{\sqrt{V_{U^\epsilon_{x}}}}.
\end{eqnarray}
Introducing a partition $\mathcal{P}$ of the 3-manifold $\Sigma$
into cells $C$, we have an operator $\hat{H}^\epsilon_{C,\gamma}$
acting on any cylindrical function $f_\gamma\in
Cyl^3_\gamma(\overline{\mathcal{A}})\otimes
Cyl_\gamma(\overline{\mathcal{U}})$ via a state-dependent
triangulation $T(\epsilon)$,
\begin{eqnarray}
\hat{H}^\epsilon_{C,\gamma}&=&\sum_{v\in
V(\gamma)}\chi_C(v)\sum_{v(\Delta)=v}\frac{\hat{p}_{\Delta}}{\sqrt{\hat{E}(v)}}\hat{h}^{\epsilon,\Delta}_{GR,v}
\frac{\hat{p}_{\Delta}}{\sqrt{\hat{E}(v)}}\nonumber\\
&+&\sum_{v\in V(\gamma)}\chi_C(v)
[\delta^{ij}(\hat{w}^{\epsilon,v}_{\gamma,\phi,i})^\dagger\hat{w}^{\epsilon,v}_{\gamma,\phi,j}
+(\hat{w}^{\epsilon,v}_{\gamma,Kin})^\dagger\hat{w}^{\epsilon,v}_{\gamma,Kin}],
\end{eqnarray}
where
\begin{eqnarray}
\hat{h}^{\epsilon,\Delta}_{GR,v}&=&\frac{16}{3i\hbar\kappa^2\gamma}
\epsilon^{ijk}\mathrm{Tr}\big(\hat{A}(\alpha_{ij}(\Delta))^{-1}\hat{A}(s_k(\Delta))^{-1}[\hat{A}(s_k(\Delta)),
\sqrt{\hat{V}_{U^\epsilon_{v}}}]\big)\nonumber\\
&+&(1+\gamma^2)\frac{8\sqrt{2}}{3i\hbar^3\kappa^4\gamma^3}\epsilon^{ijk}
\mathrm{Tr}\big(\hat{A}(s_i(\Delta))^{-1}[\hat{A}(s_i(\Delta)),\hat{K}^\epsilon]\nonumber\\
&\times&\hat{A}(s_j(\Delta))^{-1}[\hat{A}(s_j(\Delta)),\hat{K}^\epsilon]
\hat{A}(s_k(\Delta))^{-1}[\hat{A}(s_k(\Delta)),\sqrt{\hat{V}_{U^\epsilon_{v}}}]\big),\nonumber\\
\hat{w}^{\epsilon,v}_{\gamma,\phi,i}&=&\frac{i}{\hbar^2}\sum_{v(\Delta)=v}\frac{\hat{p}_{\Delta}}{\sqrt{\hat{E}(v)}}
\epsilon^{lmn}\frac{1}{\lambda}\hat{U}(v,\lambda)^{-1}[\hat{U}(t(s_l(\Delta)),\lambda)-\hat{U}(v,\lambda)]\nonumber\\
&\times&\mathrm{Tr}\big(\tau_i\hat{A}(s_m(\Delta))^{-1}[\hat{A}(s_m(\Delta)),\hat{V}_{U^\epsilon_{v}}^{5/8}]
\hat{A}(s_n(\Delta))^{-1}[\hat{A}(s_n(\Delta)),\hat{V}_{U^\epsilon_{v}}^{5/8}]\big)
\frac{\hat{p}_{\Delta}}{\sqrt{\hat{E}(v)}},\nonumber\\
\hat{w}^{\epsilon,v}_{\gamma,Kin}&=&\frac{1}{(i\hbar)^3}\sum_{v(\Delta)=v}\frac{\hat{p}_{\Delta}}{\sqrt{\hat{E}(v)}}
\hat{\pi}(v)\epsilon^{lmn}\nonumber\\
&\times&\mathrm{Tr}\big(\hat{A}(s_l(\Delta))^{-1}[\hat{A}(s_l(\Delta)),\hat{V}^{5/12}_{U^\epsilon_{v}}]
\hat{A}(s_m(\Delta))^{-1}[\hat{A}(s_m(\Delta)),\hat{V}^{5/12}_{U^\epsilon_{v}}]\nonumber\\
&\times&\hat{A}(s_n(\Delta))^{-1}[\hat{A}(s_n(\Delta)),\hat{V}^{5/12}_{U^\epsilon_{v}}]\big)
\frac{\hat{p}_{\Delta}}{\sqrt{\hat{E}(v)}}.
\end{eqnarray}
The notations here are as same as those in section 3. Note that
$\hat{H}^\epsilon_{C,\gamma}$ is similar to the Hamiltonian
constraint operator $\hat{\mathcal{H}}(1)$ defined in last section,
but is now divided by the square root of volume operator. Hence the
action of $\hat{H}^\epsilon_{C,\gamma}$ on a cylindrical function
$f_\gamma$ adds analytical arcs $a_{ij}(\Delta)$ with
$\frac{1}{2}$-representation (or arbitrary chosen spin
$j$-representation of $SU(2)$ if one uses non-fundamental
representations to express the holonomies in (\ref{GRhamilton})) and
points at $t(s_i(\Delta))$ with representation $\lambda$ with
respect to each vertex $v(\Delta)$ of $\gamma$. Thus, for each
$\epsilon>0$, $\hat{H}^\epsilon_{C,\gamma}$ is a $SU(2)$ gauge
invariant and diffeomorphism covariant operator defined on
$Cyl^3_{\gamma}(\overline{\mathcal{A}})\otimes
Cyl_{\gamma}(\overline{\mathcal{U}})$. The family of such operators
with respect to different graphs are cylindrically consistent up to
diffeomorphisms. So the inductive limit operator $\hat{H}_C$ is
densely defined on $\mathcal{H}_{Kin}$ by the uniform Rovelli-Smolin
topology. Moreover, the adjoint operators of
$\hat{H}^\epsilon_{C,\gamma}$, which are also cylindrically
consistent up to diffeomorphisms, read
\begin{eqnarray}
(\hat{H}^\epsilon_{C,\gamma})^\dagger&=&\sum_{v\in
V(\gamma)}\chi_C(v)\sum_{v(\Delta)=v}\frac{\hat{p}_{\Delta}}{\sqrt{\hat{E}(v)}}(\hat{h}^{\epsilon,\Delta}_{GR,v})^\dagger
\frac{\hat{p}_{\Delta}}{\sqrt{\hat{E}(v)}}\nonumber\\
&+&\sum_{v\in V(\gamma)}\chi_C(v)
[\delta^{ij}(\hat{w}^{\epsilon,v}_{\gamma,\phi,j})^\dagger\hat{w}^{\epsilon,v}_{\gamma,\phi,i}
+(\hat{w}^{\epsilon,v}_{\gamma,Kin})^\dagger\hat{w}^{\epsilon,v}_{\gamma,Kin}].\label{adjoint}
\end{eqnarray}
The inductive limit operator of (\ref{adjoint}) is denoted by
$(\hat{H}^\epsilon_{C})^\dagger$, which is adjoint to $\hat{H}_C$ as
\begin{eqnarray}
<g_{\gamma\ '}, \hat{H}_{C}f_{\gamma}>_{kin}&=&<g_{\gamma\
'},\hat{H}_{C,\gamma}f_{\gamma}>_{kin}=<(\hat{H}_{C,\gamma})^\dagger
g_{\gamma\ '}, f_{\gamma}>_{kin}\nonumber\\
&=&<(\hat{H}_{C})^\dagger g_{\gamma\ '},
f_{\gamma}>_{kin}=<(\hat{H}_{C})^\dagger_{\gamma\ '} g_{\gamma\ '},
f_{\gamma}>_{kin}.
\end{eqnarray}
Then a master constraint operator, $\hat{\mathbf{M}}$, in ${\cal
H}_{Diff}$ can be defined as:
\begin{equation}
(\hat{\textbf{M}}\Psi_{Diff})[\Pi_{s,c}]:=\lim_{\mathcal{P}\rightarrow
\Sigma;\epsilon,\epsilon'\rightarrow\mathrm{0}}\Psi_{Diff}[\sum_{C\in\mathcal{P}}
\frac{1}{2}\hat{H}^\epsilon_{C} (\hat{H}_{C}^{\epsilon'})^\dagger
\Pi_{s,c}].\label{master}
\end{equation}
Since $\hat{H}^\epsilon_{C} (\hat{H}^{\epsilon'}_{C})^\dagger
\Pi_{s,c}$ is a finite linear combination of spin-scalar-network
functions on an extended graph with skeleton $\gamma$, the value of
$(\hat{\textbf{M}}\Psi_{Diff})[\Pi_{s,c}]$ is finite for a given
$\Psi_{Diff}\in {\cal H}_{Diff}$. So $\hat{\textbf{M}}\Psi_{Diff}$
is in the algebraic dual of the space of cylindrical functions.
Moreover, we can show that it is diffeomorphism invariant. For any
diffeomorphism transformation $\varphi$,
\begin{eqnarray}
(\hat{U}'_\varphi\hat{\textbf{M}}\Psi_{Diff})[f_\gamma]&=&\lim_{\mathcal{P}\rightarrow
\Sigma;\epsilon,\epsilon'\rightarrow\mathrm{0}}\Psi_{Diff}[\sum_{C\in\mathcal{P}}\frac{1}{2}\hat{H}^\epsilon_{C}
(\hat{H}^{\epsilon'}_{C})^\dagger\hat{U}_\varphi
f_\gamma]\nonumber\\
&=&\lim_{\mathcal{P}\rightarrow
\Sigma;\epsilon,\epsilon'\rightarrow\mathrm{0}}\Psi_{Diff}[\hat{U}_\varphi\sum_{C\in\mathcal{P}}
\frac{1}{2}\hat{H}^{\varphi^{-1}(\epsilon)}_{\varphi^{-1}(C)}
(\hat{H}^{\varphi^{-1}(\epsilon')}_{\varphi^{-1}(C)})^\dagger f_\gamma]\nonumber\\
&=&\lim_{\mathcal{P}\rightarrow
\Sigma;\epsilon,\epsilon'\rightarrow\mathrm{0}}\Psi_{Diff}[\sum_{C\in\mathcal{P}}\frac{1}{2}\hat{H}^\epsilon_{C}
(\hat{H}^{\epsilon'}_{C})^\dagger f_\gamma],
\end{eqnarray}
for any cylindrical function $f_\gamma$, where in the last step, we
used the fact that the diffeomorphism transformation $\varphi$
leaves the partition invariant in the limit
$\mathcal{P}\rightarrow\Sigma$ and relabel $\varphi(C)$ to be $C$.
So we have the result
\begin{eqnarray}
(\hat{U}'_\varphi\hat{\textbf{M}}\Psi_{Diff})[f_\gamma]=(\hat{\textbf{M}}\Psi_{Diff})[f_\gamma].\label{diff}
\end{eqnarray}
On the other hand, given any diffeomorphism invariant
spin-scalar-network state $\Pi_{[s,c]}$, the norm of the result
state $\hat{\textbf{M}}\Pi_{[s,c]}$ can be expressed as:
\begin{eqnarray}
&&||\hat{\textbf{M}}\Pi_{[s,c]}||_{Diff}\nonumber\\
&=&\sum_{[s',c']}|<\hat{\textbf{M}}\Pi_{[s,c]}|\Pi_{[s',c']}>_{Diff}|^2 \nonumber\\
&=&\sum_{[s',c']}|\lim_{\mathcal{P}\rightarrow
\Sigma;\epsilon,\epsilon'\rightarrow\mathrm{0}}\Pi_{[s,c]}[\sum_{C\in\mathcal{P}}\frac{1}{2}\hat{H}^\epsilon_{C}
(\hat{H}^{\epsilon'}_{C})^\dagger \Pi_{s',c'\in[s',c']}]|^2\nonumber\\
&=&\sum_{[s',c']}|\lim_{\mathcal{P}\rightarrow
\Sigma;\epsilon,\epsilon'\rightarrow\mathrm{0}}\frac{1}{n_{\gamma(s,c)}}\sum_{\varphi\in
Diff/Diff_{\gamma(s,c)}}\sum_{\varphi'\in
GS_{\gamma(s,c)}}\nonumber\\
&\times&<\hat{U}_{\varphi}\hat{U}_{\varphi'}\Pi_{s,c\in[s,c]}
|\sum_{C\in\mathcal{P}}\frac{1}{2}\hat{H}^\epsilon_{C}
(\hat{H}^{\epsilon'}_{C})^\dagger \Pi_{s',c'\in[s',c']}>_{Kin}|^2\nonumber\\
&=&\sum_{[s',c']}|\lim_{\mathcal{P}\rightarrow
\Sigma;\epsilon,\epsilon'\rightarrow\mathrm{0}}\frac{1}{n_{\gamma(s,c)}}\sum_{\varphi\in
Diff/Diff_{\gamma(s,c)}}\sum_{\varphi'\in
GS_{\gamma(s,c)}}\nonumber\\
&\times&<\hat{U}_{\varphi}\hat{U}_{\varphi'}\sum_{C\in\mathcal{P}}\frac{1}{2}\hat{H}^{\epsilon'}_{C}
(\hat{H}^{\epsilon}_{C})^\dagger \Pi_{s,c\in[s,c]}
|\Pi_{s',c'\in[s',c']}>_{Kin}|^2\nonumber\\
&=&\sum_{[s',c']}|\lim_{\mathcal{P}\rightarrow
\Sigma;\epsilon,\epsilon'\rightarrow\mathrm{0}}\Pi_{[s',c']}
[\sum_{C\in\mathcal{P}}\frac{1}{2}\hat{H}^{\epsilon'}_{C}
(\hat{H}^{\epsilon}_{C})^\dagger
\Pi_{s,c\in[s,c]}]|^2,\label{dense}
\end{eqnarray}
where we make use of the fact that $\hat{\textbf{M}}$ commutes with
diffeomorphism transformations. Note that, the cylindrical function
$\sum_{C\in\mathcal{P}}\frac{1}{2}\hat{H}^{\epsilon'}_{C}
(\hat{H}^{\epsilon}_{C})^\dagger \Pi_{s,c\in[s,c]}$ is a finite
linear combination of spin-scalar-network functions on some extended
graph, so that there are only finite number of terms contributing to
the sum in Eq.(\ref{dense}). Hence it automatically converges. So
the master constraint operator $\hat{\mathbf{M}}$ defined by
Eq.(\ref{master}) is densely defined on ${\cal H}_{Diff}$.

We now compute the matrix elements of $\hat{\mathbf{M}}$. Given two
diffeomorphism invariant spin-scalar-network functions
$\Pi_{[s_1,c_1]}$ and $\Pi_{[s_2,c_2]}$, the matrix element of
$\hat{\mathbf{M}}$ is calculated as
\begin{eqnarray}
&&<\Pi_{[s_1,c_1]}|\hat{\textbf{M}}|\Pi_{[s_2,c_2]}>_{Diff}\nonumber\\
&=&\overline{(\hat{\textbf{M}}\Pi_{[s_2,c_2]})[\Pi_{s_1,c_1\in[s_1,c_1]}]}\nonumber\\
&=&\lim_{\mathcal{P}\rightarrow
\Sigma;\epsilon,\epsilon'\rightarrow\mathrm{0}}\sum_{C\in\mathcal{P}}\frac{1}{2}
\overline{\Pi_{[s_2,c_2]}[\hat{H}^\epsilon_{C}(\hat{H}^{\epsilon'}_{C})^\dagger\Pi_{s_1,c_1\in[s_1,c_1]}]}\nonumber\\
&=&\lim_{\mathcal{P}\rightarrow
\Sigma;\epsilon,\epsilon'\rightarrow\mathrm{0}}\sum_{C\in\mathcal{P}}\frac{1}{2}
\frac{1}{n_{\gamma(s_2,c_2)}}\sum_{\varphi\in
Diff/Diff_{\gamma(s_2,c_2)}}\sum_{\varphi'\in
GS_{\gamma(s_2,c_2)}}\nonumber\\
&\times&\overline{<\hat{U}_{\varphi}\hat{U}_{\varphi'}\Pi_{s_2,c_2\in[s_2,c_2]}
|\hat{H}^\epsilon_{C}(\hat{H}^{\epsilon'}_{C})^\dagger\Pi_{s_1,c_2\in[s_1,c_1]}>_{Kin}}\nonumber\\
&=&\sum_{s,c}\lim_{\mathcal{P}\rightarrow
\Sigma;\epsilon,\epsilon'\rightarrow\mathrm{0}}\sum_{C\in\mathcal{P}}\frac{1}{2}
\frac{1}{n_{\gamma(s_2,c_2)}}\sum_{\varphi\in
Diff/Diff_{\gamma(s_2,c_2)}}\sum_{\varphi'\in
GS_{\gamma(s_2,c_2)}}\nonumber\\
&\times&\overline{<\hat{U}_{\varphi}\hat{U}_{\varphi'}\Pi_{s_2,c_2\in[s_2,c_2]}
|\hat{H}^\epsilon_{C,\gamma(s,c)}\Pi_{s,c}>_{Kin}<\Pi_{s,c}|(\hat{H}^{\epsilon'}_{C})^\dagger\Pi_{s_1,c_1\in[s_1,c_1]}>_{Kin}}\nonumber\\
&=&\sum_{[s,c]}\sum_{v\in
V(\gamma(s,c\in[s,c]))}\frac{1}{2}\lim_{\epsilon,\epsilon'\rightarrow\mathrm{0}}\nonumber\\
&\times&\overline{\Pi_{[s_2,c_2]}[\hat{H}^{\epsilon}_{v,\gamma(s,c)}
\Pi_{s,c\in[s,c]}]\sum_{s,c\in[s,c]}
<\Pi_{s,c}|(\hat{H}^{\epsilon'}_{v})^\dagger\Pi_{s_1,c_1\in[s_1,c_1]}>_{Kin}},\label{matrix}
\end{eqnarray}
where $Diff_{\gamma}$ is the set of diffeomorphisms leaving the
colored graph $\gamma$ invariant, $GS_{\gamma}$ denotes the graph
symmetry quotient group $Diff_{\gamma}/TDiff_{\gamma}$ where
$TDiff_{\gamma}$ is the diffeomorphisms which is trivial on the
graph $\gamma$, and $n_\gamma$ is the number of elements in
$GS_\gamma$. Note that we have used the resolution of identity trick
in the fourth step. Since only finite number of terms in the sum
over spin-scalar-networks $(s,c)$, cells $C\in\mathcal{P}$, and
diffeomorphism transformations $\varphi$ are non-zero respectively,
we can interchange the sums and the limit. In the fifth step, we
take the limit $C\rightarrow v$ and split the sum $\sum_{s,c}$ into
$\sum_{[s,c]}\sum_{s,c\in[s,c]}$, where $[s,c]$ denotes the
diffeomorphism equivalent class associated with $(s,c)$. Here we
also use the fact that, given $\gamma(s,c)$ and $\gamma(s',c')$
which are different up to a diffeomorphism transformation, there is
always a diffeomorphism $\varphi$ transforming the graph associated
with $\hat{H}^{\epsilon}_{v,\gamma(s,c)} \Pi_{s,c}\
(v\in\gamma(s,c))$ to that of $\hat{H}^{\epsilon}_{v'\gamma(s',c')}
\Pi_{s',c'}\ (v'\in\gamma(s',c'))$ with $\varphi(v)=v'$, hence
$\Pi_{[s_2,c_2]}[\hat{H}^{\epsilon}_{v,\gamma(s,c)}
\Pi_{s,c\in[s,c]}]$ is constant for different $(s,c)\in[s,c]$.

Since the term $\sum_{s,c\in[s,c]}
<\Pi_{s,c}|(\hat{H}^{\epsilon'}_{v})^\dagger\Pi_{s_1,c_1\in[s_1,c_1]}>_{Kin}$
is independent of the parameter $\epsilon'$, one can see that by
fixing an arbitrary state-dependent triangulation $T(\epsilon')$,
\begin{eqnarray}
&&\sum_{s,c\in[s,c]}
<\Pi_{s,c}|(\hat{H}^{\epsilon'}_{v})^\dagger\Pi_{s_1,c_1\in[s_1,c_1]}>_{Kin}\nonumber\\
&=&\sum_{\varphi}<U_\varphi\Pi_{s,c}|(\hat{H}^{\epsilon'}_{v})^\dagger\Pi_{s_1,c_1\in[s_1,c_1]}>_{Kin}\nonumber\\
&=&\sum_{\varphi}<\hat{H}^{\epsilon'}_{v,\varphi(\gamma(s,c))}U_\varphi\Pi_{s,c}|\Pi_{s_1,c_1\in[s_1,c_1]}>_{Kin}\nonumber\\
&=&\sum_{\varphi}<U_\varphi\hat{H}^{\varphi^{-1}(\epsilon')}_{\varphi^{-1}(v),\gamma(s,c)}\Pi_{s,c}|\Pi_{s_1,c_1\in[s_1,c_1]}>_{Kin}\nonumber\\
&=&\overline{\Pi_{[s_1,c_1]}[\hat{H}^{\varphi^{-1}(\epsilon')}_{v\in
V(\gamma(s,c)),\gamma(s,c)}\Pi_{s,c}]},
\end{eqnarray}
where $\varphi$ are the diffeomorphism transformations spanning the
diffeomorphism equivalent class $[s,c]$. Note that the kinematical
inner product in above sum is non-vanishing if and only if
$\varphi(\gamma(s,c)))$ coincides with the extended graph obtained
from certain skeleton $\gamma(s_1,c_1)$ by the action of
$(\hat{H}^{\epsilon'}_{v})^\dagger$ and $v\in
V(\varphi(\gamma(s,c)))$, i.e., the scale $\varphi^{-1}(\epsilon')$
of the diffeomorphism images of the tetrahedrons added by the action
coincides with the scale of certain tetrahedrons in $\gamma(s,c)$
and $\varphi^{-1}(v)$ is a vertex in $\gamma(s,c)$. Then we can
express the matrix elements (\ref{matrix}) as:
\begin{eqnarray}
&&<\Pi_{[s_1,c_1]}|\hat{\textbf{M}}|\Pi_{[s_2,c_2]}>_{Diff}\nonumber\\
&=&\sum_{[s,c]}\sum_{v\in
V(\gamma(s,c\in[s,c]))}\frac{1}{2}\lim_{\epsilon,\epsilon'\rightarrow\mathrm{0}}
\overline{\Pi_{[s_2,c_2]}[\hat{H}^{\epsilon}_{v,\gamma(s,c)}
\Pi_{s,c\in[s,c]}]}\Pi_{[s_1,c_1]}[\hat{H}^{\epsilon'}_{v,\gamma(s,c)}\Pi_{s,c\in[s,c]}]\nonumber\\
&=&\sum_{[s,c]}\sum_{v\in
V(\gamma(s,c\in[s,c]))}\frac{1}{2}\overline{(\hat{H}'_v\Pi_{[s_2,c_2]})[
\Pi_{s,c\in[s,c]}]}(\hat{H}'_v\Pi_{[s_1,c_1]}) [
\Pi_{s,c\in[s,c]}].\label{master2}
\end{eqnarray}
From Eq.(\ref{master2}) and the result that the master constraint
operator $\hat{\mathbf{M}}$ is densely defined on
$\mathcal{H}_{Diff}$, it is obvious that $\hat{\mathbf{M}}$ is a
positive and symmetric operator on ${\cal H}_{Diff}$. Hence, it is
associated with a unique self-adjoint operator
$\hat{\overline{\mathbf{M}}}$, called the Friedrichs extension of
$\hat{\mathbf{M}}$. We relabel $\hat{\overline{\mathbf{M}}}$ to be
$\hat{\mathbf{M}}$ for simplicity. In conclusion, there exists a
positive and self-adjoint operator $\hat{\mathbf{M}}$ on
$\mathcal{H}_{Diff}$ corresponding to the master constraint
(\ref{mconstraint}). It is then possible to obtain the physical
Hilbert space of the coupled system by the direct integral
decomposition of $\mathcal{H}_{Diff}$ with respect to
$\hat{\mathbf{M}}$.

Note that the quantum constraint algebra can be easily checked to be
anomaly free. Eq.(\ref{diff}) assures that the master constraint
operator commutes with finite diffeomorphism transformations, i.e.,
\begin{eqnarray}
[\hat{\mathbf{M}},\hat{U}'_\varphi]=0.
\end{eqnarray}
Also it is obvious that the master constraint operator commutes with
itself,
\begin{eqnarray}
[\hat{\mathbf{M}},\hat{\mathbf{M}}]=0.
\end{eqnarray}
So the quantum constraint algebra is precisely consistent with the
classical constraint algebra (\ref{malgebra}) in this sense. As a
result, the difficulty of the original Hamiltonian constraint
algebra can be avoided by introducing the master constraint algebra,
due to the Lie algebra structure of the latter.

\section*{Acknowledgments}

The authors would like to acknowledge the referee for helpful
criticism on the original manuscript. This work is supported in part
by NSFC (10205002). Muxin Han would also like to acknowledge the
support from the fellowship and the assistantship of LSU, Hearne
Foundation of LSU, and funding from Advanced Research and
Development Activity.

\end{document}